# Domain wall conduction in multiaxial ferroelectrics

Eugene A. Eliseev[1,2], Anna N. Morozovska[1*], George S. Svechnikov[1],
Peter Maksymovych[3] and Sergei V. Kalinin[3†]

[1] Institute of Semiconductor Physics, National Academy of Science of Ukraine,
41, pr. Nauki, 03028 Kiev, Ukraine

[2] Institute for Problems of Materials Science, National Academy of Science of Ukraine,
3, Krjijanovskogo, 03142 Kiev, Ukraine

[3] Center for Nanophase Materials Science, Oak Ridge National Laboratory,
Oak Ridge, TN, 37831

**Abstract**

The conductance of domain wall structures consisting of either stripes or cylindrical domains in multi-axial ferroelectric-semiconductors is analyzed. The effects of the domain size, wall tilt and curvature, on charge accumulation, are analyzed using the Landau-Ginsburg Devonshire (LGD) theory for polarization combined with Poisson equation for charge distributions. Both the classical ferroelectric parameters including expansion coefficients in 2-4-6 Landau potential and gradient terms, as well as flexoelectric coupling, inhomogeneous elastic strains and electrostriction are included in the present analysis. Spatial distributions of the ionized donors, free electrons and holes were found self-consistently using the effective mass approximation for the respective densities of states. The proximity and size effect of the electron and donor accumulation/depletion by thin stripe domains and cylindrical nanodomains are revealed. In contrast to thick domain stripes and thicker cylindrical domains, in which the carrier accumulation (and so the static conductivity) sharply increases at the domain walls only, small nanodomains of radius less then 5-10 correlation length appeared conducting across entire cross-section. Implications of such conductive nanosized channels may be promising for nanoelectronics.

[*] morozo@i.com.ua

[†] sergei2@ornl.gov

# 1. Introduction

Ferroelectric domain walls were recently shown to act as conductive channels in ferroelectric-dielectrics and ferroelectric-semiconductors even at room-temperature, providing experimental counterparts to decade-old theoretical predictions [1]. The experimental results in materials such as BiFeO$_3$ [2, 3]. Pb(Zr,Ti)O$_3$ [4], SbSJ [5] and LiNbO$_3$ doped with MgO [6], all enabled by the development of scanning probe microscopy techniques capable of probing the conductance on the nanoscale, suggest the universality of this behavior. These results present an obvious interest for fundamental studies of ferroic and low-dimensional systems physics, as well as offer a potential pathway for oxide nanoelectronics due to their nanosized width as well as the possibility to control their spatial location by external fields [4]. However, for a given ferroelectric material, the wall conductivity should depend on the wall tilt, local strains (due to electrostriction), and proximity effects. These factors in turn determine the possibility for multilevel storage, device size, and integration into solid-state devices. Thus the understanding of the role of these effects on wall conductivity is a required first step in analyzing the feasibility of controllable rewritable conductive nanosized channels design in otherwise non-conductive ferroelectrics.

## I. Historical overview

We summarize data on wall conductance in uniaxial materials, wall structure in multiaxial materials, and mechanisms of coupling between order parameters and strain, which are relevant to the analysis of wall conductance.

### I.1. Wall conductance in uniaxial materials

Recent reviews of up-to-date theoretical achievements in the field of domain structures in ferroics could be found in many textbooks (see e.g. [7, 8]). Briefly, the consistent studies of ferroelectric domain wall (DW) begins with seminal papers of Zhirnov [9] and Cao&Cross [10], who considered 180- and 90-degree DW taking into account electrostriction coupling between the spontaneous polarization and strain, but considering only electro-neutral DW. The case of rhombohedral symmetry is considered in Ref. [11]. Note, that orientation of 180 degree DW is determined by electrostatics, while orientation of 90-degree twin DW is mainly governed by the strain compatibility [9, 10, 12].

Earlier results on domains in uniaxial ferroelectric semiconductors are summarized in Ref. [13], recent studies [14] and [15] are devoted to the perpendicular (or "counter") and inclined DW respectively. The static conductivity of domain walls with different incline angle with respect to spontaneous polarization vector in the *uniaxial* ferroelectrics-semiconductors of *n*-type was calculated numerically [15]. Unexpectedly, the static conductivity drastically increases at the inclined head-to-head wall by 1 order of magnitude for small incline angles up 3 orders of magnitude for the perpendicular domain wall due to the strong electron accumulation.

At the same time the study of DW structure and conductance in multiaxial ferroelectrics is much more complicated, since there are several components of order parameter, which should be mixed at the DW through strain, biquadratic coupling term and flexoelectric effect as discussed below.

### I.2. Wall structure in multiaxial materials

For multiaxial ferroics with multicomponent order parameters, analysis of polarization structure at the domain wall necessitates taking into account the coupling between the order parameter components components (e.g. for boundary between 90-degree DW or some type of 180-degree in incipient ferroelectrics [16]), mediated by stress accommodation or gradients coupling. For instance, the bi-quadratic in two order parameters coupling, also known as Houchmandzadeh-Lajzerowicz-Salje coupling [17], was introduced to describe the coupling between the polarization and structural order parameter (see Ref.[18] for typical case of PZT). This coupling can lead to the appearance of polarization on the structural domains (twins), however the conditions of such manifestations are usually very strict [16]. Situation is similar for ferromagnetic-ferroelectric, where local magnetic moment is possible at the ferroelectric DW due to either biquadratic [19] or inhomogeneous coupling [20, 21].

Despite the fact that attempts to describe polarization behavior in multicomponent ferroics were attempted since the early days of ferroelectricity [9, 10, 22], the progress with understanding of their DW structure appeared very limited. Only recently Hlinka and Márton [23] calculated numerically the structure of twin boundaries in tetragonal perovskite crystal $BaTiO_3$ in the framework of the phenomenological LGD model. They found that the polarization component normal to DW demonstrates a weak deviation from constant distribution, in contrast to the previous studies of Zhirnov [9], and Cao & Cross [10]. This leads to the internal electric field appearance and hence to the potential step at DW, which is consistent with *ab initio* calculations [24]. Ferroelectric DW resembling Neel walls in ferromagnetics was predicted in thin ferroelectric films [25] and incipient ferroelectrics 26, 27].

### I.3. Flexoelectric effect on wall structure

It should be noted that none of the previous theoretical studies predict normal component of polarization at the nominally neutral 180-degree domain walls in the bulk ferroelectrics. At the same time the flexoelectric coupling can break the wall symmetry and can induce the normal component of polarization along 180-degree DW [4]. Flexoelectric effect describes the coupling of polarization with strain gradient and polarization gradient with the strain [28, 29, 30], and was predicted by Mashkevich and Tolpygo [31]. Subsequently, a number of theoretical studies of the flexoelectric effect in conventional [32, 33, 34, 35, 36, 37, 38, 39] and incipient [40] ferroelectrics have been performed. Experimental measurements of flexoelectric tensor components were recently carried out by Ma and Cross [41, 42, 43] and Zubko et al [44]. Recently very high value of flexoelectric coupling coefficient was reported [45] for a polar phase of polyvinylidene fluoride films.

It is generally believed that the main consequence of the flexoelectric coupling is the renormalization of the polarization gradient energy (see e.g. [16, 32, 38]). In addition, some unusual coupling terms originated from the flexoelectric effect in nanosystems [38, 46]. Notably, the flexoelectric coupling could not be ignored in the presence of inhomogeneous strains/stress, and hence becomes relevant in the vicinity of the surfaces/interfaces and domain walls. For example, the presence in the free energy expansion of invariants such as $P_1 \partial(P_3^2)/\partial x_1$ allowed even in the isotropic solid leads to the existence of the (generally overlooked) polarization component normal to the nominally neutral DW. Hence polarization distributions, related to "neutral" 180 degree DW ($P_3(x_1)$) leads the appearance of internal electric field proportional to the polarization gradient, $\partial(P_3^2)/\partial x_1$.

Here we explore the polarization structure and transport behavior at the domain walls in the **multiaxial** ferroelectrics like BiFeO$_3$ and Pb(Zr,Ti)O$_3$ determined by the interplay between the polarization components and strong flexoelectric coupling between polarization components and inhomogeneous elastic strains along the walls. Original part of the paper is organized as following. Basic equations are listed and discussed in the Section 2. The impact of the tilt angle and flexoelectric coupling on the polarization vector, potential, electric field and carrier redistribution across the stripe domains is analyzed in the Section 3.1. The impact of the proximity and finite size effect and flexoelectric coupling on the polarization vector, potential, electric field and carrier redistribution across the thin stripes and cylindrical nanodomains is analyzed in the Sections 3.2 and 3.3 correspondingly. Section 4 is a brief summary.

**2. Basic equations**

Here we analyze the charge accumulation by various ferroelectric domain structures using LGD formalism. The bulk free energy density is:

$$G[\mathbf{P}, \hat{X}] = \Delta G_b + \Delta G_{elast} + \Delta G_{strict} + \Delta G_{flexo} - P_i \frac{E_i^d}{2} + \frac{g_{ijkl}}{2} \frac{\partial P_i}{\partial x_j} \frac{\partial P_k}{\partial x_l} \qquad (1)$$

**P** is ferroelectric polarization vector and **X** is the stress tensor. The tensor of gradients coefficients $g_{ijkl}$ is positively defined for commensurate ferroelectrics considered hereinafter. $E_i^d = -\frac{\partial \varphi}{\partial x_i}$ is the depolarization field components, caused by imperfect screening by the surrounding and inhomogeneous polarization distribution and/or its breaks at interfaces.

The polarization-dependent density $\Delta G_b$ for the parent phase of m3m symmetry (e.g. for tetragonal, orthorhombic, and rhombohedral ferroic phases) considered hereinafter can be written as a Taylor series expansion of the polarization, $P_i$ ($i$=1-3), as [10]:

$$\Delta G_b = a_1\left(P_1^2 + P_2^2 + P_3^2\right) + a_{11}\left(P_1^4 + P_2^4 + P_3^4\right) + a_{12}\left(P_1^2 P_2^2 + P_2^2 P_3^2 + P_3^2 P_1^2\right) + a_{111}\left(P_1^6 + P_2^6 + P_3^6\right) \\ + a_{112}\left[P_1^4\left(P_2^2 + P_3^2\right) + P_2^4\left(P_3^2 + P_1^2\right) + P_3^4\left(P_1^2 + P_2^2\right)\right] + a_{123}\left(P_1^2 P_2^2 P_3^2\right) \quad (2)$$

Here $a_i$, $a_{ij}$ and $a_{ijk}$ are the dielectric stiffness and higher-order stiffness coefficients at constant stress. The elastic energy in Eq.(1) is

$$\Delta G_{elast} = -\frac{1}{2}s_{11}\left(X_1^2 + X_2^2 + X_3^2\right) - s_{12}\left(X_1 X_2 + X_2 X_3 + X_3 X_1\right) - \frac{1}{2}s_{44}\left(X_4^2 + X_5^2 + X_6^2\right) \quad (3)$$

Here the stress components are $X_i$ ($i$=1-6 in Voigt notation).

The coupling energy between polarization and strain $\Delta G_{strict}$ is proportional to electrostriction coefficients:

$$\Delta G_{strict} = \begin{pmatrix} -Q_{11}\left(X_1 P_1^2 + X_2 P_2^2 + X_3 P_3^2\right) - Q_{44}\left(X_4 P_2 P_3 + X_5 P_3 P_1 + X_6 P_1 P_2\right) \\ -Q_{12}\left(X_1\left(P_2^2 + P_3^2\right) + X_2\left(P_3^2 + P_1^2\right) + X_3\left(P_1^2 + P_2^2\right)\right) \end{pmatrix} \quad (4)$$

where $s_{ij}$ are the elastic compliance coefficient at constant polarization; $Q_{ij}$ is the electrostriction tensor.

Flexoelectric coupling contribution is

$$\Delta G_{flexo} = -\frac{f_{ikl}}{2}\left(X_i\frac{\partial P_k}{\partial x_l} - \frac{\partial X_i}{\partial x_k}P_l\right) \quad (5)$$

Flexoelectric effect tensor is denoted as $f_{ikl}$. Full form of Eq.(5) is rather cumbersome and given by (A.1) in the **Appendix A**.

The electrostatic potential, φ satisfies the Poisson equation

$$\varepsilon_0 \varepsilon_b \Delta\varphi = \text{div}(\mathbf{P}) - e\left(N_d^+(\varphi) + p(\varphi) - n(\varphi) - N_a^-\right) \quad (6)$$

Here $\Delta$ is the Laplace operator, the charges are in the units of electron charge $e=1.6\times10^{-19}$ C, $\varepsilon_0=8.85\times10^{-12}$ F/m is the universal dielectric constant, $\varepsilon_b$ is the background dielectric permittivity of the material (unrelated with the soft mode), that is typically much smaller than the ferroelectric permittivity $\varepsilon_{ij}^f$ related with the soft mode. Note that the ferroelectric permittivity is already included in Eq.(6) from the term $\text{div}(\mathbf{P})$, when ferroelectric polarization can be approximated as expansion $P_i = P_i^S + \varepsilon_{ij}^f E_j + ...$.

Ionized deep acceptors with field-independent concentration $N_a^-$ play the role of a background charge. The equilibrium concentrations of ionized shallow donors $N_d^+$ (e.g. vacancies), free electrons $n$ and holes $p$ are:

$$N_d^+(\varphi) = N_{d0}(1 - f(E_d - E_F - e\varphi)), \tag{7a}$$

$$p(\varphi) = \int_0^\infty d\varepsilon \cdot g_p(\varepsilon) f(\varepsilon - E_V + E_F + e\varphi), \tag{7b}$$

$$n(\varphi) = \int_0^\infty d\varepsilon \cdot g_n(\varepsilon) f(\varepsilon + E_C - E_F - e\varphi). \tag{7c}$$

Where $N_{d0}$ is the donors concentration, $f(x) = \{1 + \exp(x/k_B T)\}^{-1}$ is the Fermi-Dirac distribution function, $k_B = 1.3807 \times 10^{-23}$ J/K, $T$ is the absolute temperature. $E_F$ is the Fermi energy level, $E_d$ is the donor level, $E_C$ is the bottom of conductive band, $E_V$ is the top of the valence band (all energies are defined with respect to the vacuum level). In the effective mass approximation the density of states are $g_n(\varepsilon) \approx \dfrac{\sqrt{2 m_n^3 \varepsilon}}{2\pi^2 \hbar^3}$ and $g_p(\varepsilon) \approx \dfrac{\sqrt{2 m_p^3 \varepsilon}}{2\pi^2 \hbar^3}$. Typically the condition $m_n \ll m_p$ is satisfied.

Assuming that a single domain bulk ferroelectric is electroneutral at electric potential $\varphi=0$, the condition should be valid $N_a^- = N_{d0}^+ + p_0 - n_0$, where for $\varphi=0$ the equilibrium concentrations are

$$N_{d0}^+ = N_{d0}(1 - f(E_d - E_F)) \equiv N_{d0} f(E_F - E_d), \qquad p_0 = \int_0^\infty d\varepsilon \cdot g_p(\varepsilon) f(\varepsilon + E_F - F_V) \qquad \text{and}$$

$$n_0 = \int_0^\infty d\varepsilon \cdot g_n(\varepsilon) f(\varepsilon + E_C - F_F).$$

Since the quasi-one dimensional distribution of polarization and stresses in the vicinity of the domain walls will depend only on the distance from the wall plane, it is convenient to switch from the axes $x_1$, $x_2$, $x_3$ along the cubic symmetry axes to new coordinate system, with $\tilde{x}_1$ axis normal to the domain wall plane $\{\tilde{x}_2, \tilde{x}_3\}$ [**Figs.1a**]. Rotations of crystallographic reference frame to the coordinate system, associated with the domain wall, are defined by the angles $\{\theta, \phi\}$. Components of any vector (e.g. polarization, field) and tensor (e.g. stress) in new coordinate system could be written as $\tilde{P}_i = A_{ij} P_j$, $\tilde{E}_i^d = A_{ij} E_j^d$ and $\tilde{X}_{ij} = A_{ik} A_{jl} X_{kl}$, where

$$\mathbf{A} = \begin{pmatrix} \cos\theta\cos\phi & \cos\theta\sin\phi & \sin\theta \\ -\sin\phi & \cos\phi & 0 \\ -\sin\theta\cos\phi & -\sin\theta\sin\phi & \cos\theta \end{pmatrix} \tag{8}$$

Inverse relations are $P_i = A_{ij}^T \tilde{P}_j$, $E_i^d = A_{ij}^T \tilde{E}_j^d$ and $X_{ij} = A_{ik}^T A_{jl}^T \tilde{X}_{kl}$, where is $\mathbf{A^T}$ the inverse (i.e. transposed) matrix $\mathbf{A}$. Contribution of the inhomogeneous strains $\tilde{u}_i$ to the free energy could be evaluated as:

$$\tilde{F} = G[A_{ij}^T \tilde{P}_j, A_{ij}^T \tilde{X}_j, A_{ij}^T \tilde{E}_j^d] + \tilde{X}_i \tilde{u}_i \tag{9}$$

Other contributions to the free energy could be obtained by the coordinate system transformation. Corresponding equations of state are $\frac{\partial \widetilde{F}}{\partial \widetilde{P}_i} = 0$ and $\frac{\partial \widetilde{F}}{\partial \widetilde{X}_i} = 0$.

Additional constraints on the system are given by mechanical equilibrium conditions, $\partial X_{ij}(\mathbf{x})/\partial x_i = 0$, and compatibility relations, $e_{ikl} e_{jmn} \left( \partial^2 u_{ln}/\partial x_k \partial x_m \right) = 0$, where $e_{ikl}$ is the permutation symbol or anti-symmetric Levi-Civita tensor [47]. Finally, mechanical boundary conditions for zero stress at mechanically free surfaces (e.g. $X_{3j}(z=0)=0$) should be satisfied.

Boundary conditions are determined by the configuration of the domain structure in a straightforward way. In particular, the potential vanishes ($\varphi(\widetilde{x}_1 \to \pm\infty)=0$) far from the plane of a single 180-degree domain wall and reaches maximum at the wall, so the depolarization field component $\widetilde{E}_1$ normal to the domain wall plane is zero at the wall ($\widetilde{E}_1(\widetilde{x}_1=0)=0$). The polarization components $\widetilde{P}_i$ are zero at the wall plane.

In the next sections we analyze the cases of a tilted domain stripes **[Figs.1b]**, parallel domain stripes **[Figs.1c]** and a single cylindrical domain **[Figs.1d]** regarding one-dimensional distribution of polarization in the vicinity of the domain walls.

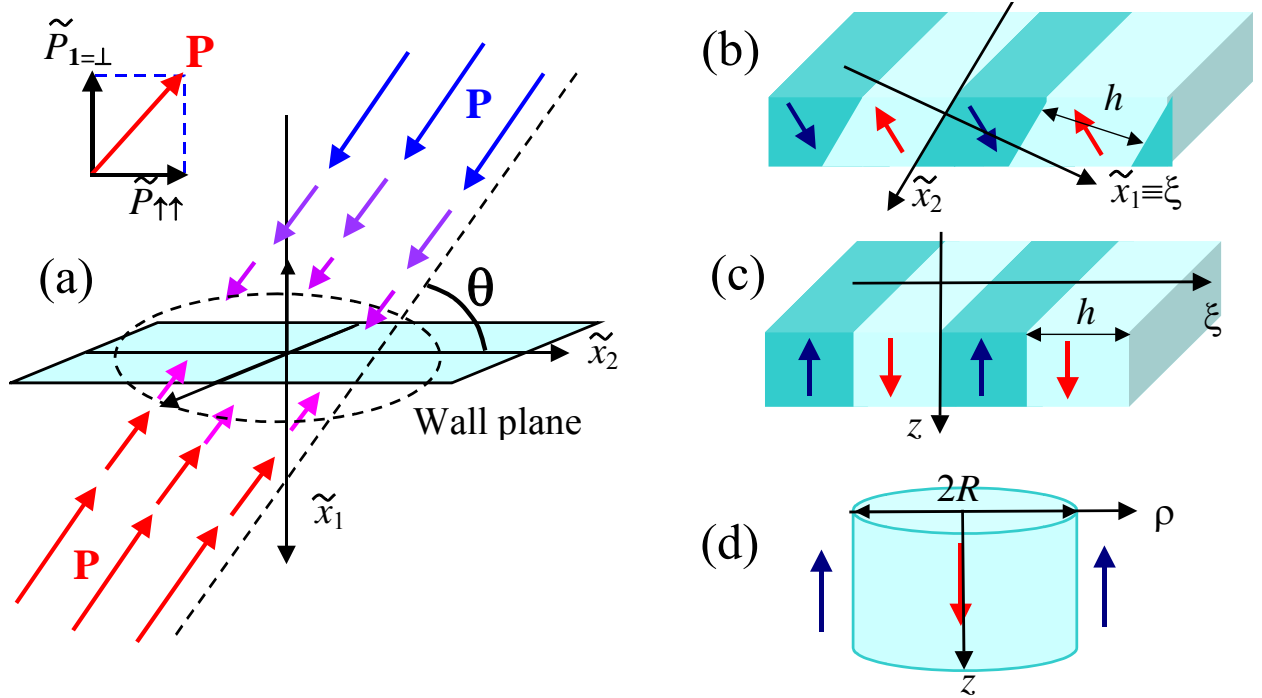

**Fig. 1.** Quasi-one dimensional distribution of polarization in the vicinity of a single domain wall (a), tilted (b) and parallel (c) domain stripes with half-period $h$; (d) cylindrical domain of radius $R$. Arrows in plots b-d indicate the polarization direction in the center of domains.

We note that for a charged domain wall the zig-zag instabilities at the wall plane can develop to minimize electrostatic energy [48]. However, here we consider only the cases for which the quasi-one dimensional distribution of polarization (periodic domain stripes, cylindrical domains, etc.) and leave the question of wall stability to further studies.

### 3. Results and discussion
### 3.1. Carriers accumulation on 180-degree domain stripes

Here we consider the effect of the wall tilt and flexoelectric coupling on the carriers redistribution in a stripe domain structure, consisting of the thin 180-degree domains of half-period $h$ that is much higher than a correlation length, $r_c = \sqrt{-g_{44}/2a_1}$. The planes $\xi = nh$ ($n = 0,\pm 1,\pm 2,...$) correspond to the domain wall between two neighboring stripes (see **Fig.1b**). In the section we regard that $h = 100 r_c$. The condition $h \gg r_c$ allows us to concentrate on the impact of the wall tilt and flexoelectric coupling, while proximity effects, which are dominant for thin stripes, will be considered in the next section.

Equation of state $\dfrac{\partial \widetilde{F}}{\partial \widetilde{P}_i} = 0$ and $\dfrac{\partial \widetilde{F}}{\partial \widetilde{X}_i} = 0$ and the Poisson equation (6) were transformed in dimensionless variables (see **Appendix A.2**) and then analyzed numerically. Results of numerical modeling for the polarization components, electric field and carriers redistribution across the domain wall between the neighboring stripes are shown in **Figs. 2-4**. PbZr$_{0.2}$Ti$_{0.8}$O$_3$ (**PZT**) material parameters used in the calculations are listed in the **Table B1** in **Appendix B**. Correlation length $r_c \approx 0.5$ nm, coordinate $\xi \equiv \widetilde{x}_1$, spontaneous polarization $P_S$ and thermodynamic coercive field $E_{coer}$ are introduced. Estimations based on Ma and Cross [49] results give the flexoelectric effect coefficients $\left|f_{ijkl}\right| \approx (0.5 - 1) \times 10^{-10}$ m$^3$/C.

Dependencies of the polarization components perpendicular ($\widetilde{P}_1(\xi) \equiv P_\perp(\xi)$) and parallel ($\widetilde{P}_{\uparrow\uparrow}(\xi)$) to the wall plane, electric potential ($\varphi(\xi)$) and field ($\widetilde{E}_1(\xi) \equiv \widetilde{E}_\perp(\xi)$), concentrations of electrons, ionized donors on the distance $\xi$ from the wall plane between the neighboring stripes was calculated for the **tilted** domain stripes with different tilt angles $\theta = \pi/2, \pi/4, \pi/6, \pi/30, 0$ (see curves 1, 2, 3, 4, 5 in **Figs.2-3**). Plots in **Fig. 2** are calculated **without flexoelectric coupling**, when only electrostriction couples polarization and elastic strains. Plots in **Fig. 3** and **Fig. 4** are calculated **with positive and negative flexoelectric coupling coefficient** $f_{12}$ respectively.

Unexpectedly, the wall thickness between the neighboring stripes appeared different with respect to the components $P_\perp(\xi)$ and $\widetilde{P}_{\uparrow\uparrow}(\xi)$. In particular, the parallel wall ($\theta = 0$) is the thickest for $P_\perp(\xi)$ and the thinnest for $\widetilde{P}_{\uparrow\uparrow}(\xi)$. The perpendicular wall with maximal bound charge ($\theta = \pi/2$) is the thinnest for $P_\perp(\xi)$ and the thickest for $\tilde{P}_{\uparrow\uparrow}(\xi)$ (**Fig.2a, 3a** and **4a**). This suggests that the electrostriction coupling leads only to the narrowing of the domain walls.

Electro-physical properties of the domain walls between the neighboring stripes are shown in **Figs.2-4**. Note, that head-to-head (**h-t-h**) and tail-to-tail (**t-t-t**) domain walls reveal very different electronic properties: head-to-head walls appeared electron accumulating (see electron accumulating regions at $\xi \approx \pm 100 r_c$ in **Fig.2e, 3e**) and thus have high conductivity in the considered n-type ferroelectric-semiconductors, while tail-to-tail walls appeared donor (e.g. vacancies) accumulating (see donor accumulating region $|\xi| < 10 r_c$ in **Fig.2f, 3f**), similarly to the one-component polarization in uniaxial ferroelectrics considered in [15].

The electric field and potential created by the wall bound charges and screening carriers are the highest for the perpendicular wall ($\theta = \pi/2$) with maximal bound charge $2P_S$. Since the bound charge is $2P_S \sin\theta$, it decreases with the with $\theta$ decrease, and vanishes at $\theta = 0$ (**Fig.2c,d** and **3c,d**). Electrostriction induces local increase of the coercive field at the domain wall between the neighboring stripes (see **Fig.2d, 3d** and **4d**, where the maximal field is higher that thermodynamic coercive one $E_{coer}$). The electron concentration is the highest for the head-to-head perpendicular wall ($\theta = \pi/2$); it decreases with the bound charge decrease (i.e. with $\theta$ decrease) and vanishes at $\theta = 0$ (compare maximal values for different curves in **Fig.2e** and **3e**). The electron accumulation leads to the strong increase of the static conductivity across the charged domain stripes in PZT up 3 orders of magnitude for the perpendicular domain walls. Holes concentration appeared less than $10^{-40}$m$^{-3}$, i.e. free holes are almost absent near the head-to-head domain walls between the neighboring stripes.

The noticeable differences between **Figs.2c-f** and **3c-f, 4c-f** originated from the flexoelectric effect via the coupling coefficient $f_{12}$. At zero or small tilt angle an additional features on the field, potential, electrons and ionized donor concentration appear in vicinity of domain walls due to the nonzero flexoelectric coupling (compare curves in **Figs.2c-f** with the curves in Figs. **3c-f, 4c-f**). Note that the flexoelectric and tilt effects are not additive. The flexoeffect-induced feature position is independent on wall tilt angle, but its width increases with the angle decrease. This indicates that the flexoelectric coupling leads to the appearance of additional potential well/barrier on the electrostatic potential distribution depending on the flexoelectric coefficient $f_{12}$ sign (see **Fig.3c** and **Fig.4c**).

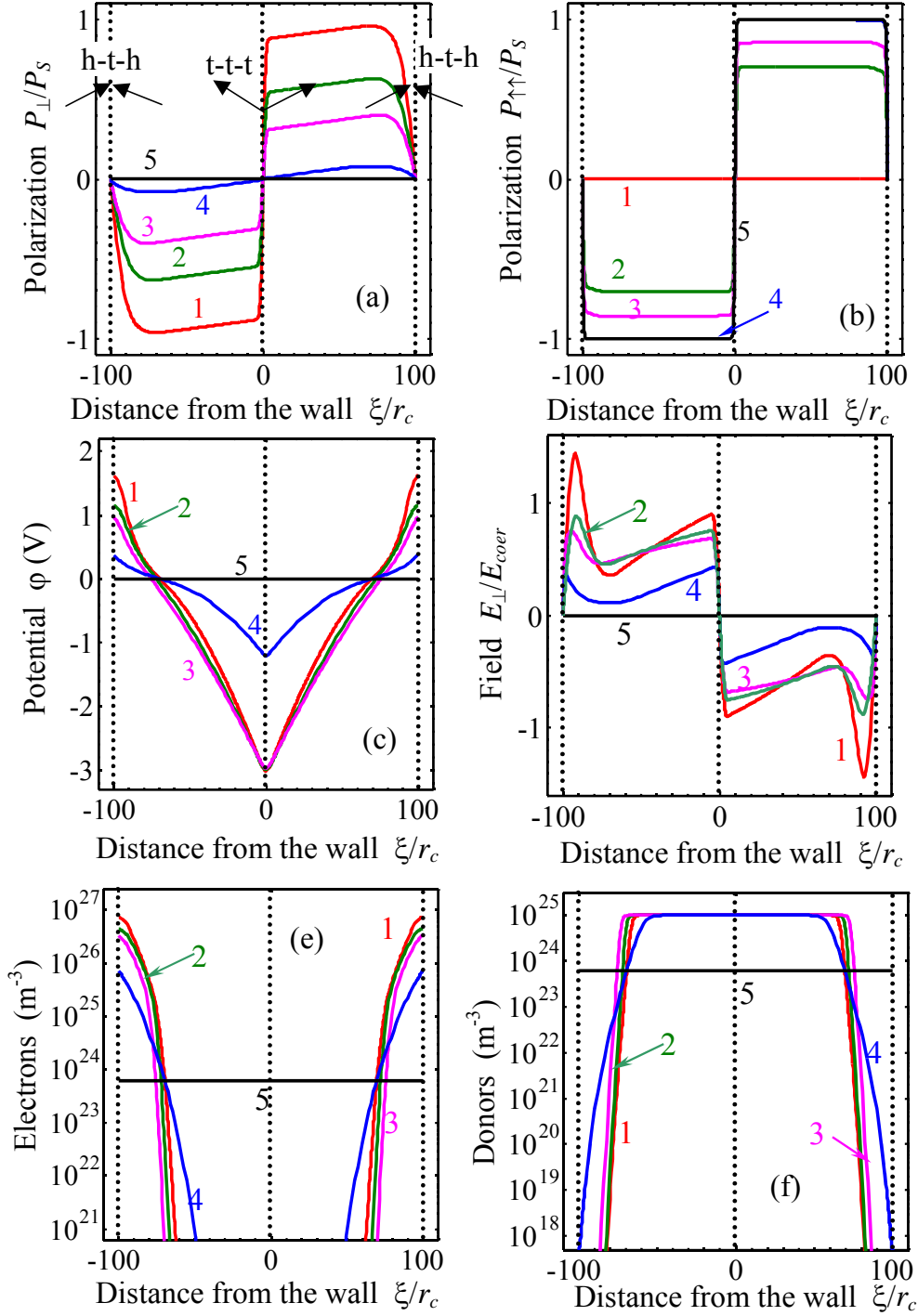

**Fig. 2.** Dependencies of the polarization component $\widetilde{P}_\perp(\xi)/P_S$ perpendicular to the wall plane (a) and (b) the component $\widetilde{P}_{\uparrow\uparrow}(\xi)/P_S$ parallel to the wall plane, potential $\varphi(\xi)$ (c), field $\widetilde{E}_\perp(\xi)/E_{coer}$ (d), concentrations of electrons $n(\xi)$ (e) and ionized donors $N_d^+(\xi)$ (f) calculated **without flexoelectric coupling** for the tilted 180-degree domain wall between the neighboring stripes with different tilt angles $\theta = \pi/2, \pi/4, \pi/6, \pi/30, 0$ (curves 1, 2, 3, 4, 5) and **zero flexoelectric coupling** $f_{ijkl} = 0$. Material parameters correspond to PbTi$_{0.8}$Zr$_{0.2}$O$_3$ (listed in the **Table B1** in **Appendix B**), $N_{d0} = 10^{25}$ m$^{-3}$, $E_d = 0.1$ eV, stripe half-period $h = 100 r_c$.

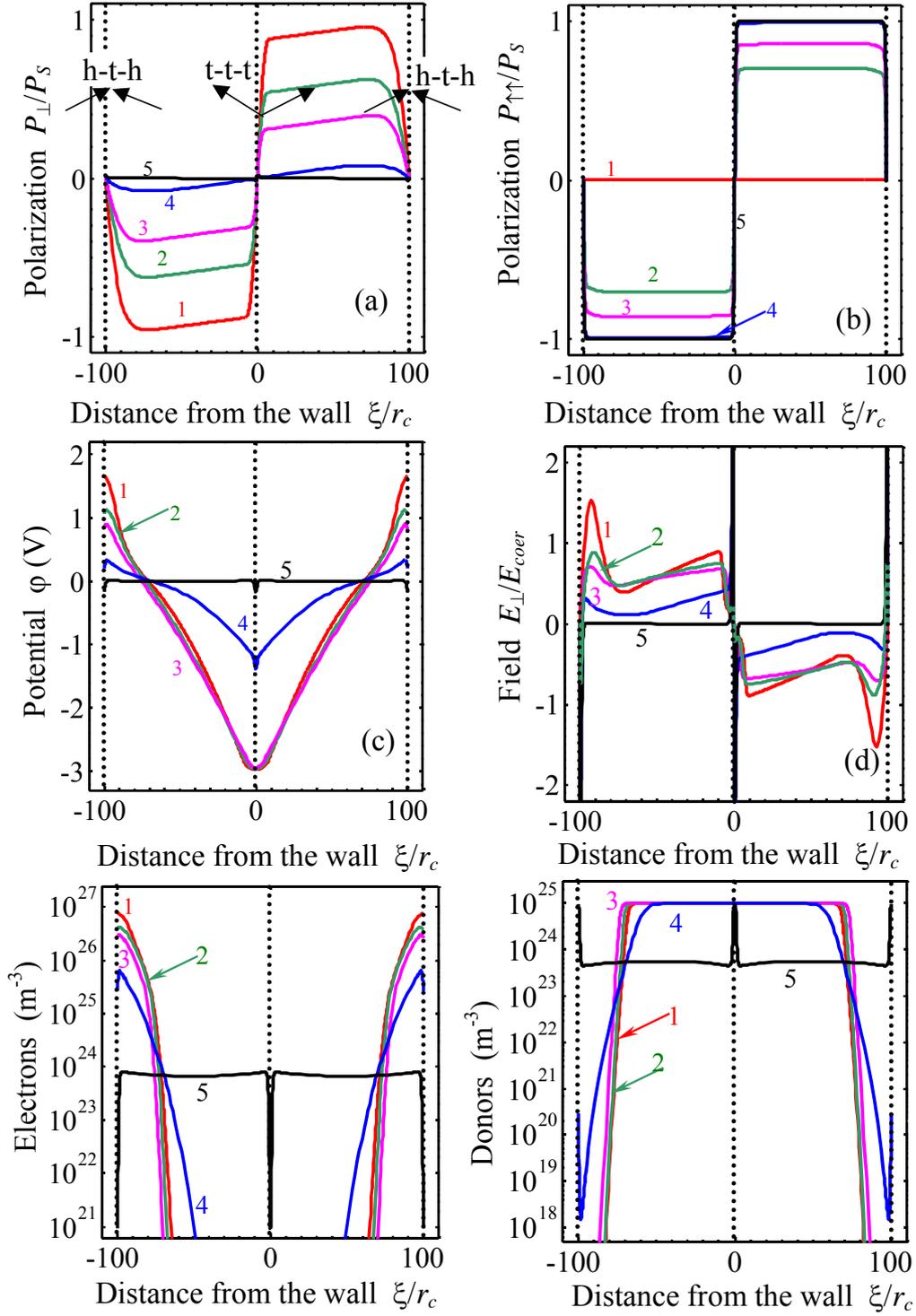

**Fig. 3.** Dependencies of the polarization component $\tilde{P}_\perp(\xi)/P_S$ perpendicular to the wall plane (a) and (b) the component $\tilde{P}_{\uparrow\uparrow}(\xi)/P_S$ parallel to the wall plane, potential $\varphi(\xi)$ (c), field $\tilde{E}_\perp(\xi)/E_{coer}$ (d), concentrations of electrons $n(\xi)$ (e) and ionized donors $N_d^+(\xi)$ (f) calculated for the tilted 180-degree domain wall between the neighboring stripes with different tilt angles $\theta = \pi/2, \pi/4, \pi/6, \pi/30, 0$ (curves 1, 2, 3, 4, 5) and **positive flexoelectric coupling** ($f_{12} = +0.5 \times 10^{-10}$ m$^3$/C). Parameters are the same as in **Figs.2.**

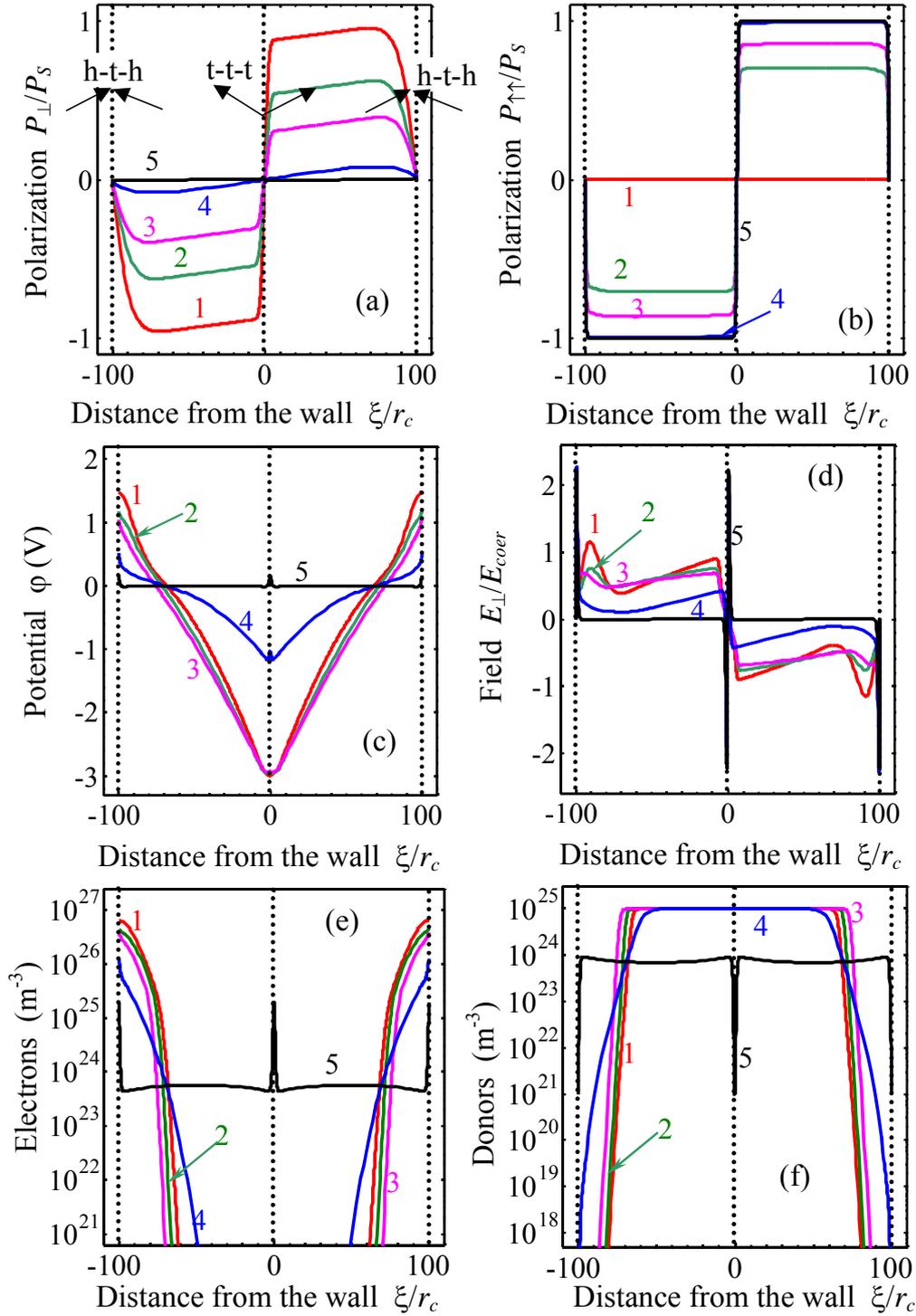

**Fig. 4.** Dependencies of the polarization component $\tilde{P}_\perp(\xi)/P_S$ perpendicular to the wall plane (a) and (b) the component $\tilde{P}_{\uparrow\uparrow}(\xi)/P_S$ parallel to the wall plane, potential $\varphi(\xi)$ (c), field $\tilde{E}_\perp(\xi)/E_{coer}$ (d), concentrations of electrons $n(\xi)$, (e) and ionized donors, (f) calculated for the tilted $180^0$-domain wall between the neighboring stripes with different tilt angles $\theta = \pi/2, \pi/4, \pi/6, \pi/30, 0$ (curves 1, 2, 3, 4, 5) and **negative flexoelectric coupling** ($f_{12} = -0.5 \times 10^{-10}$ m$^3$/C). Material parameters are the same as in **Figs.2.**

To the best of our knowledge, the effect of the flexoelectric coupling on the ferroelectric wall charge state was not studied theoretically before. At the same time, flexoelectric coupling leads to the nontrivial physical responses, including appearance of $P_\perp(\xi)$ and its strong gradient across the "nominally uncharged" 180-degree domain wall. At the same time, polarization component $P_{\uparrow\uparrow}(\xi)$ is only weakly affected by the presence of the flexoelectric coupling.

We note that the flexoelectric coupling term $f_{12}\tilde{P}_1 \partial(\tilde{X}_2 + \tilde{X}_3)/\partial\tilde{x}_1$ in the free energy causes the effective "flexoelectric" field $f_{12} \partial(\tilde{X}_2 + \tilde{X}_3)/\partial\tilde{x}_1$ acting on $P_\perp(\xi)$ component due to the partial mechanical clamping of the wall. The field leads to the carriers re-distribution and thus to conductivity changes even across the nominally uncharged parallel domain walls due the appearance of the polarization component perpendicular to the wall plane [4]. At that the depth of the potential barrier well/height appeared at the wall due to $P_\perp(\xi)$-effect, as derived in **Appendix D,** is proportional to the flexoelectric coupling coefficient $f_{12}$.

Note, that the static electronic and ionic conductivity can be estimated as $\sigma_e(\xi) = e(\eta_e n(\xi) + \eta_p p(\xi))$ and $\sigma_i(\xi) = e\eta_d N_d^+(\xi)$, where $\eta_{e,p,d}$ are corresponding mobilities, which are regarded constant. Since the strength of the carrier accumulation/depletion at the wall plane is determined by the electric potential $\varphi(\xi)$ behavior at the wall, the conductivity should be controlled by the field effect, which distribution across the wall in turn depends on the wall tilt, stripe domain size, etc. **Fig. 5a** shows the potential $\varphi(\xi)$ dependence on the head-to-head (h-t-h) and tail-to-tail (t-t-t) wall tilt angle θ. The potential modulation depth increases with θ increase. Dependence of the electronic and ionic conductivity at the wall between the neighboring domain stripes on the wall tilt is shown in **Fig. 5b,c** for negative, zero, and positive flexoelectric coupling coefficient. It is seen that the head-to-head wall electronic conductivity increases with tilt θ increase; head-to-head wall ionic conductivity decreases with θ increase. The tail-to-tail wall electronic conductivity decreases with tilt θ increase; tail-to-tail wall ionic conductivity increases and saturates with θ increase.

To summarize, the free carrier accumulation leads to the strong increase of the static conductivity across the tilted walls between stripe domains in multi-axial ferroelectrics-semiconductors of n-type: from 1 order for the parallel domain stripes due to the flexoelectric coupling up 3 orders of magnitude for perpendicular domain walls (even without flexoelectric coupling impact).

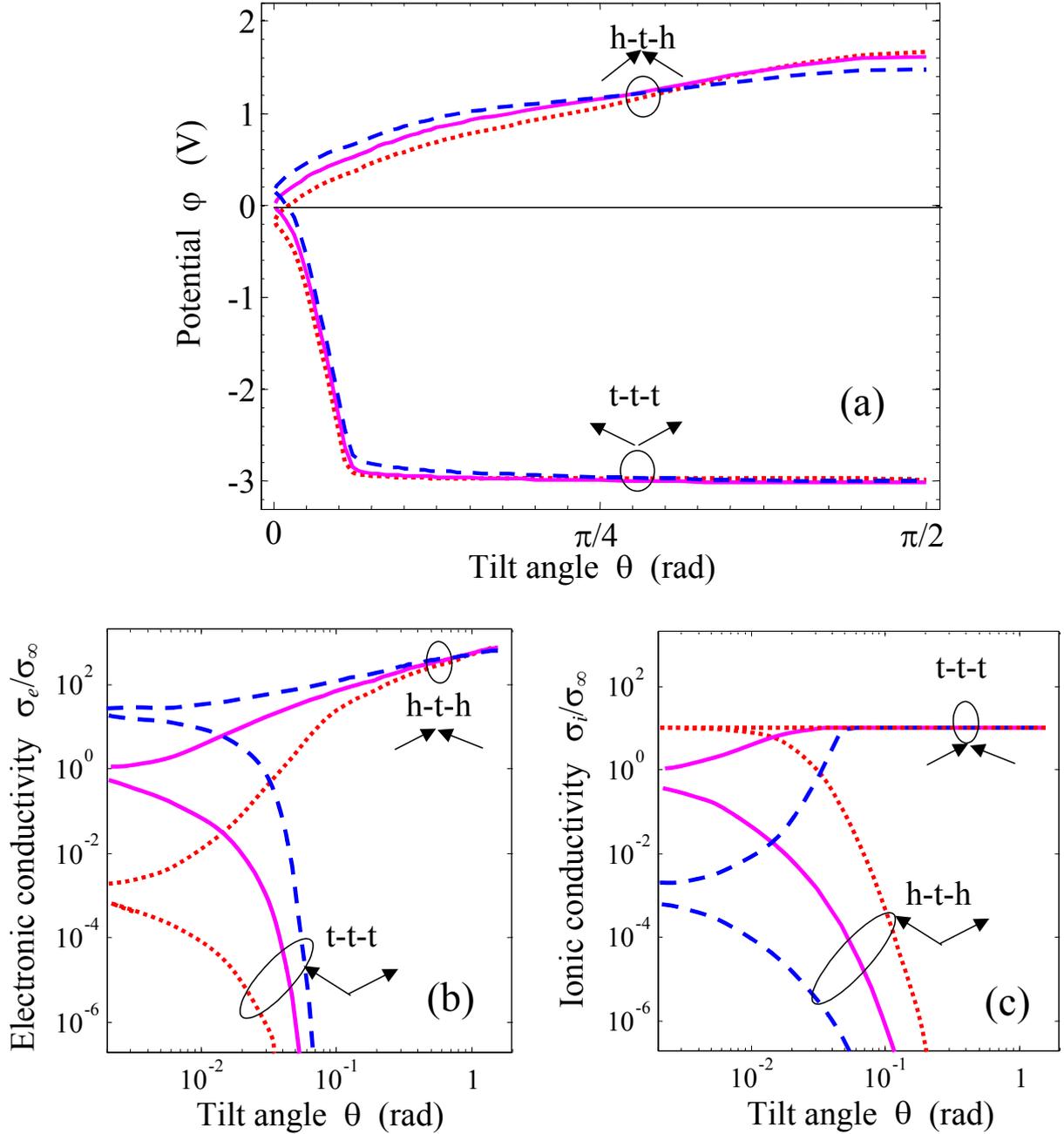

**Fig. 5.** Dependence of the potential $\varphi(\xi = 0)$ (a), electronic (b) and ionic (c) conductivity at the 180-degree domain wall between the neighboring head-to-head stripes on the head-to-head wall tilt angle $\theta$ calculated for negative, zero, and positive **flexoelectric coupling coefficient** $f_{12} = (-0.5, 0, 0.5) \times 10^{-10}$ m$^3$/C (dotted, solid and dashed curves respectively). Material parameters are the same as in **Figs.2.**

### 3.2. Proximity effects of carrier accumulation by 180-degree stripe domains

Here we consider the case when domain walls in the thin 180-degree domains of half-period $h$ with "nominally neutral" parallel domain walls ($\theta=0$) are close, i.e. proximity effects on static conductivity. Nominally neutral means that the polarization vector is parallel to the wall plane in the center of the domain stripe. The planes $\xi = nh$ ($n = 0, \pm 1, \pm 2, ...$) correspond to the domain wall between two neighboring stripes (see **Fig.1c**). Our calculations show that the polarization component $\widetilde{P}_{\perp}(\xi)$ and depolarization field $\widetilde{E}_{\perp}(\xi)$ are induced due to the flexoelectric coupling. The bound charge $\widetilde{P}_{\perp}(\xi)$ leads to the appearance of lateral electric field $\widetilde{E}_{\perp}(\xi)$ and carrier concentration redistribution in the vicinity of domain walls.

The distributions of polarization components $\widetilde{P}_{\uparrow\uparrow}(\xi)$ and $\widetilde{P}_{\perp}(\xi)$, depolarization electric field $\widetilde{E}_{\perp}(\xi)$, electrostatic potential $\varphi(\xi)$ and screening charges (electrons and donors) are shown in **Figs. 6-7** for several periods of domain stripes (different curves), and for positive (**Fig.6**) and negative (**Fig.7**) flexoelectric coefficient $f_{12}$. Note, that the stripe domains with half-period below minimal value $h_{cr} \sim 2r_c$ is thermodynamically unstable due to **proximity effect** making domain wall energy too high, so the curves in **Fig. 6-7** are plotted for the values $h \geq h_{cr}$.

$\widetilde{P}_{\perp}(\xi)$ and $\widetilde{E}_{\perp}(\xi)$ are maximal in the vicinity of domain walls (i.e. at $\xi = nh \pm \sqrt{2}r_c$), zero at the walls and in the center of domain stripe (i.e. at $\xi = nh$ and $\xi = nh \pm h/2$). The maximal value of polarization component is

$$\widetilde{P}_{\perp}^{\max} \approx \pm f_{12}\varepsilon_0\varepsilon_b \frac{(Q_{11}+Q_{12})P_S^2}{(s_{11}+s_{12})\sqrt{2}r_c}\left(1-\frac{h_{cr}}{h}\right) \tag{10}$$

and electric field

$$\widetilde{E}_{\perp}^{\max} \approx \mp f_{12}\varepsilon_0\varepsilon_b \frac{(Q_{11}+Q_{12})P_S^2}{(s_{11}+s_{12})\sqrt{2}r_c}\left(1-\frac{h_{cr}}{h}\right) \tag{11}$$

are reached in the points $\xi = nh \pm \sqrt{2}r_c$ ($n = 0,\pm 1,\pm 2,...$) corresponding to the distance $\sqrt{2}r_c$ from the domain wall planes.

Here, $h_{cr}$ is the aforementioned minimal half-period of the stable domain stripe (corresponding to the critical size originating from the **proximity effect**), which is related to correlation length $r_c$ as $h_{cr} \approx \pi r_c/2$ for $f_{12} = 0$ and θ=0. Electric potential reaches the maximal value

$$\varphi_{\max} \approx f_{12}\frac{(Q_{11}+Q_{12})P_S^2}{(s_{11}+s_{12})}\left(1-\frac{h_{cr}}{h}\right) \tag{12}$$

at the wall locations, $\xi = nh$. Note, that the expressions for $\widetilde{P}_{\perp}^{\max}$, $\widetilde{E}_{\perp}^{\max}$ and $\varphi_{\max}$ differ from the expressions listed in Ref.[4] by the factor $\left(1-\frac{h_{cr}}{h}\right)$, representing the proximity effect. As anticipated, $\widetilde{P}_{\uparrow\uparrow}(\xi)$ is maximal in the center of the domain stripes $\xi = nh \pm h/2$ and zero at the walls $\xi = nh$. Electrons and donors distributions have sharp extremum (minimum or maximum depending on the sign of $f_{12}$) at the walls $\xi = nh$.

It is seen that the decrease of the domain stripe half-period *h* leads to the gradual suppression of the both polarization components maximum as well as to the decrease of the modulation depth of the potential and screening charges profiles across the domain stripes. Polarization, potential, field and carries concentration profiles have sinusoidal shape for thin stripes. Anharmonicity appears and strongly increases with *h* increase.

The type of accumulated carriers (electrons or donors for considered case) is determined by the sigh of the flexoelectric coefficient: negative $f_{12}$ leads to the accumulation of negative carriers (electrons or acceptors), positive $f_{12}$ leads to the accumulation of positive carriers (holes, donors or vacancies) along the wall plane (see **Fig.6e,f** and **Fig.7e,f**). The higher is the $f_{12}$ value the stronger is the carrier accumulation effect. Note, that the experimental results [41, 42, 43] show that the coefficient $f_{12}$ is negative for PZT.

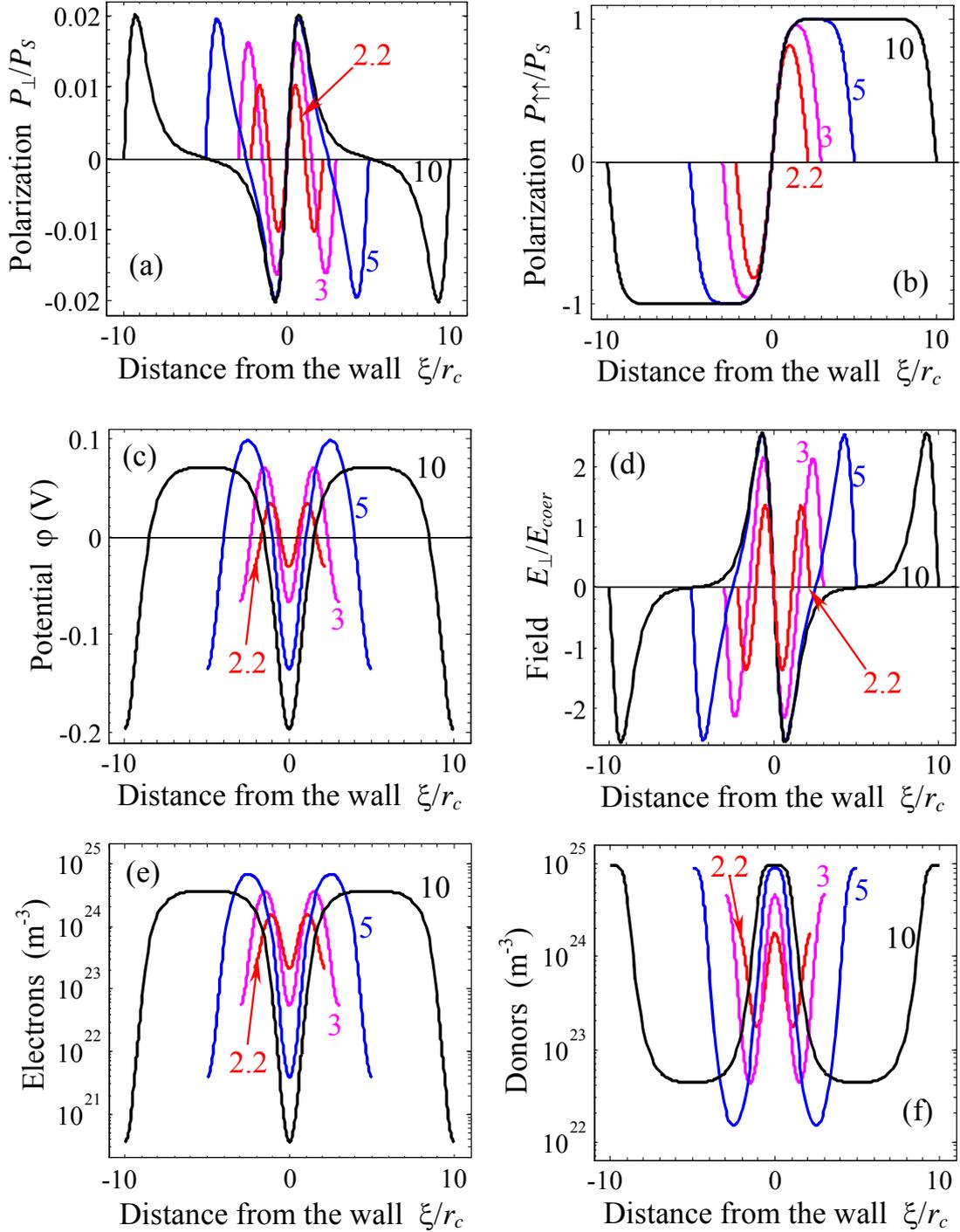

**Fig. 6.** Dependencies of the polarization component $\tilde{P}_\perp(\xi)/P_S$ perpendicular to the wall plane (a) and (b) the component $\tilde{P}_{\uparrow\uparrow}(\xi)/P_S$ parallel to the wall plane, potential $\varphi(\xi)$ (c), field $\tilde{E}_\perp(\xi)/E_{coer}$ (d), concentrations of electrons $n(\xi)$ (e) and ionized donors $N_d^+(\xi)$ (f) across the "nominally **uncharged**" 180-degree domain stripes (only one period is shown) with different half periods $h/r_c$=2.2, 3, 5, 10 (listed near the curves) and **positive flexoelectric coupling coefficient** $f_{12} = 1\times10^{-10}$m$^3$/C. Other parameters are same as for **Fig. 2**.

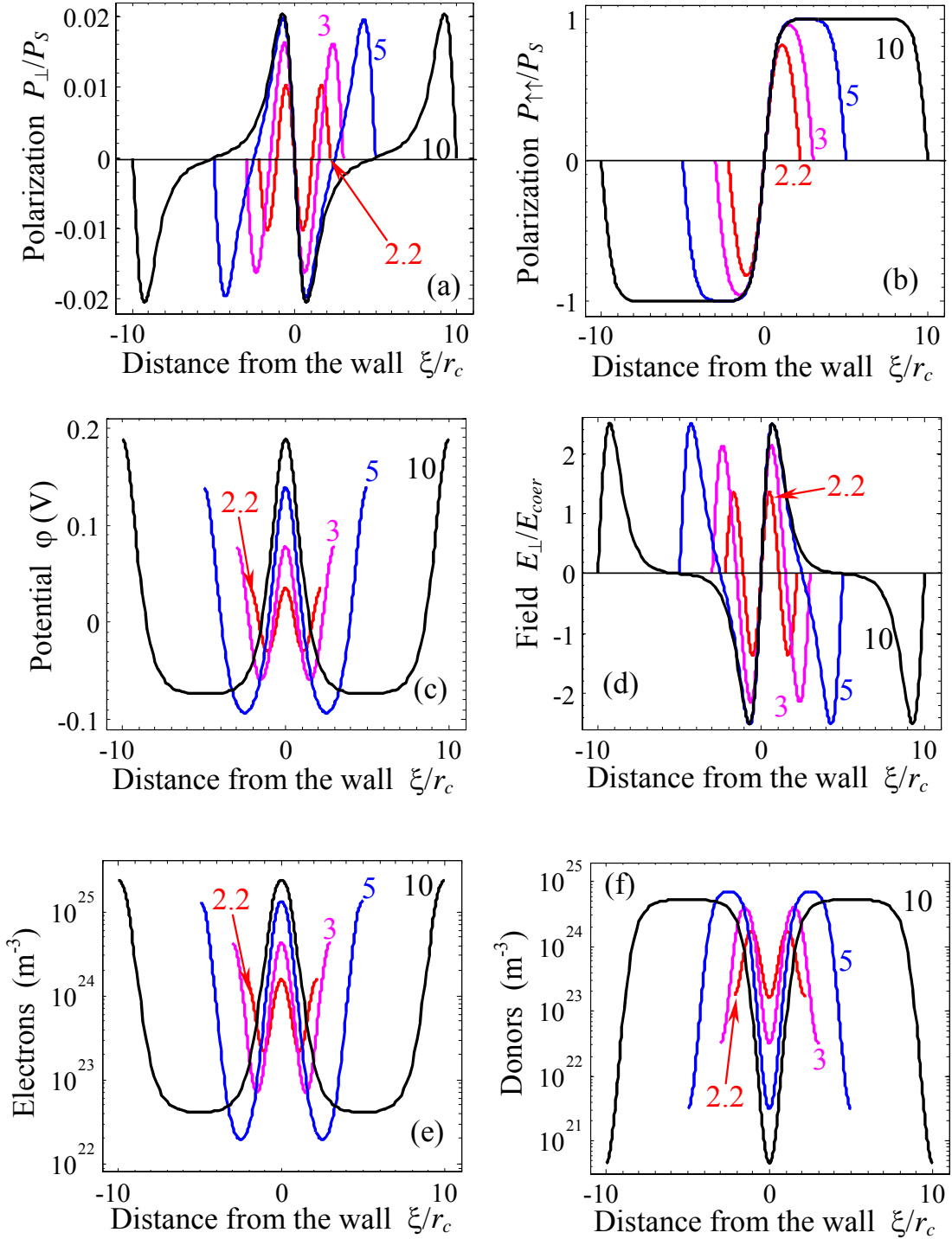

**Fig. 7.** Dependencies of the polarization component $\tilde{P}_\perp(\xi)/P_S$ perpendicular to the wall plane (a) and (b) the component $\tilde{P}_{\uparrow\uparrow}(\xi)/P_S$ parallel to the wall plane, potential $\varphi(\xi)$ (c), field $\tilde{E}_\perp(\xi)/E_{coer}$ (d), concentrations of electrons $n(\xi)$ (e) and ionized donors $N_d^+(\xi)$ (f) across the "nominally **uncharged**" 180-degree domain stripes (only one period is shown) with different half periods $h/r_c$=2.2, 3, 5, 10 (listed near the curves) and **negative flexoelectric coupling coefficient** $f_{12} = -1 \times 10^{-10}$ m$^3$/C. Other parameters are same as for **Fig. 2**.

The possibility of the electron and donor accumulation/depletion in the vicinity of the stripes domain boundaries is demonstrated in **Fig. 8.** To estimate the observable conductivity, local concentrations of electrons $n(\xi)$ and donors $N_d^+(\xi)$ should be averaged across over wall **apparent thickness**, e.g. for distance $\xi \in \{-r_c, r_c\}$ (solid curves) as well as **entire** the domain stripe $\xi \in \{-h, h\}$ (dashed curves).

Note, that $\langle n(\xi) \rangle / n(\infty) \approx \langle \sigma_e(\xi) \rangle / \sigma_e(\infty)$ and $\langle N_d^+(\xi) \rangle / N_d^+(\infty) \approx \langle \sigma_i(\xi) \rangle / \sigma_i(\infty)$ in the framework of the model adopted here. It is seen from the solid curves in **Fig. 8a** that the wall electronic conductivity monotonically increases and then saturates (up to 30 times in saturation in comparison with a bulk electronic conductivity $\sigma_e(\infty)$) with the domain stripe period increase for **negative** flexoelectric coupling. It is seen from the solid curves in **Fig. 8b** that the wall ionic conductivity monotonically increases and then saturates (up to 15 times in saturation in comparison with a bulk ionic conductivity $\sigma_i(\infty)$) with the stripe period increase for **positive** flexoelectric coupling. Without flexoelectric coupling the conductivity is the same as for the homogeneous monodoman region $\sigma_{e,i}(\infty)$ (horizontal lines "0").

Unexpectedly, the averaging entire the domain stripe smears the impact of flexoelectric coupling sign: the dashed curves are relatively close for positive $f_{12} = +1 \times 10^{-10}$ m³/C and negative $f_{12} = -1 \times 10^{-10}$ m³/C, in contract to very different solid curves. Independently on the $f_{12}$ sign conductivity averaged entire the domain stripe firstly increase with the stripe half-period $h$ increase for very small half-periods $h_{cr} < h < 5r_c$, then reaches a diffuse maximum (~ 5 times in comparison with homogeneous $\sigma_{e,i}(\infty)$) and then decreases with further $h$ increase. The principal difference in the solid and dashed curves behaviour can be explained by the following considerations. For negative $f_{12}$ free electrons accumulate in the immediate vicinity of the domain walls, the central regions of the stripes are depleted with electrons (see **Fig.7e**). For positive $f_{12}$ the immediate vicinity of the domain walls are depleted with electrons, the central regions of the stripes accumulate electrons (see **Fig.6e**). The situation with ionized donors is visa versa: ionized donors accumulation takes place in the vicinity of domain walls for positive $f_{12}$, the central regions of the stripes are depleted with donors (see **Fig.6f** and **7f**). The averaging entire the domain stripe $\xi \in \{-h, h\}$ contains information only about resulting depletion + accumulation effect. As anticipated, the total charge of "electrons + ionized donors" is exactly zero (i.e. the sum of the dashed curves "+1" or "-1") due to the total electroneutrality in the domain structure.

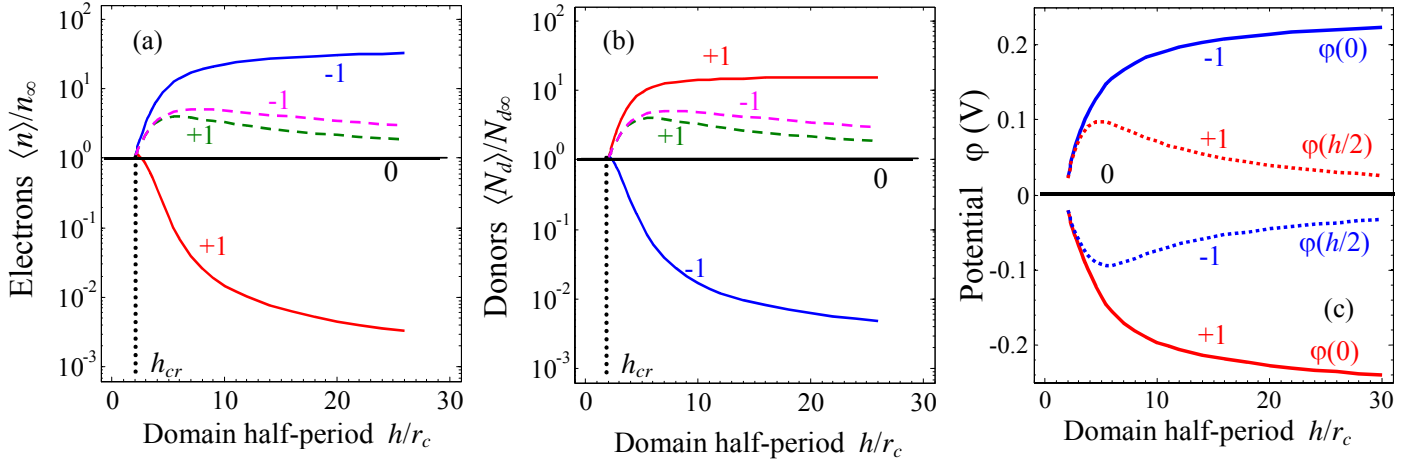

**Fig. 8.** Dependence of the relative electron $\langle n(\xi)\rangle/n(\infty)$ (a) and donor $\langle N_d^+(\xi)\rangle/N_d^+(\infty)$ (b) concentrations, potential φ (c) on the domain stripes half-period $h$ calculated for **negative** ($f_{12} = -1\times 10^{-10}$m³/C), **zero** and **positive** ($f_{12} = -1\times 10^{-10}$m³/C) **flexoelectric coupling coefficient** (numbers near the curves). Concentrations of electrons $n(\rho)$ and donors $N_d^+(\rho)$ were averaged across the range $\xi \in \{-r_c, r_c,\}$ (solid curves) as well as entire the domain cross-section $\xi \in \{-h, h\}$ (dashed curves). Potential barrier (c) is plotted at the domain wall ($\varphi(0)$, solid curves) and in the middle of the stripe ($\varphi(h/2)$, dotted curves). Material parameters are the same as in **Figs.2.**

Due to the flexoelectric coupling, even in average the domain stripes with period $h \sim 5r_c$ appeared up to 5 times more conductive then the monodomain region with $\sigma_{e,i}(\infty)$. The carriers accumulation in the domain wall region is caused by the potential barrier $\varphi(\xi)$ in turn caused by the uncompensated bound charge $P_\perp(\xi)$. Corresponding potential barrier is plotted in **Fig. 8c** at the domain wall ($\varphi(0)$, solid curves) and in the middle of the stripe ($\varphi(\pm h/2)$, dotted curves) for positive, zero and negative flexoelectric coupling coefficients. Potential barrier $\varphi(0)$ monotonically increases with the stripe size $h$ increase and then saturates. Potential $\varphi(\pm h/2)$ in the middle of the stripes firstly increases with h increase, reaches maximum at $h \sim 5r_c$ and then decreases with further $h$ increase. For thick stripes with half-period $h \gg 100r_c$ the potential vanishes in the central region of each stripe, i.e. $\varphi(|\xi| \ll h) \to 0$, as anticipated.

### 3.3. Carrier accumulation at the cylindrical domain wall

Cylindrical domain walls always form at the initial stages of local polarization reversal caused by a charged SPM probe [50, 51] or as a first stage of polarization switching in thin films. Here we consider the finite size effect of carrier accumulation and static conductivity of radially-symmetric **cylindrical domain wall** with curvature radius $R$ (see **Fig.1d**). Polar radius $\rho = \sqrt{x^2 + y^2}$ is introduced. We assume that cylinder axis is pointed along one of the possible directions of spontaneous polarization. Note that for other orientations the problem could not be considered as quasi-one dimensional. Furthermore, only the case of small radii $R \leq 10 r_c$ is of interest, since for larger radii the behavior is very similar to those obtained in the section 3.1 for the thick stripe domains.

The numerical analysis shows that the polarization component $\tilde{P}_\perp(\rho)$ and depolarization field $\tilde{E}_\perp(\rho)$ are induced due to the flexoelectric coupling. The bound charge $\tilde{P}_\perp(\rho)$ leads to the electric field $\tilde{E}_\perp(\rho)$ appearance and then to carrier concentration redistribution across the cylindrical wall. The distributions of polarization components $\tilde{P}_{\uparrow\uparrow}(\rho)$ and $\tilde{P}_\perp(\rho)$, electric field $\tilde{E}_\perp(\rho)$, electrostatic potential $\varphi(\rho)$ and screening charges (electrons and donors) are shown in **Figs. 9-10** for different domain radius (different curves), positive (**Fig.9**) and negative (**Fig.10**) flexoelectric coefficient $f_{12}$. Note, that the cylindrical domain with half-period below critical value $R_{cr} \sim 1.2 r_c$ is thermodynamically unstable due to **finite size effect** making the domain wall energy too high, so the curves in **Fig. 9-10** are plotted for the values $R \geq R_{cr}$.

$\tilde{P}_\perp(\rho)$ and $\tilde{E}_\perp(\rho)$ are maximal in the vicinity of domain walls (i.e. at $\rho = R - \sqrt{2} r_c$), zero at the walls and in the center of cylindrical domain. The maximal value of polarization component is

$$\tilde{P}_\perp^{max} \approx \pm f_{12} \varepsilon_0 \varepsilon_b \frac{(Q_{11} + Q_{12}) P_S^2}{(s_{11} + s_{12}) \sqrt{2} r_c} \left(1 - \frac{R_{cr}}{R}\right) \tag{13}$$

and electric field

$$\tilde{E}_\perp^{max} \approx \mp f_{12} \varepsilon_0 \varepsilon_b \frac{(Q_{11} + Q_{12}) P_S^2}{(s_{11} + s_{12}) \sqrt{2} r_c} \left(1 - \frac{R_{cr}}{R}\right) \tag{14}$$

are reached for $\rho = R - \sqrt{2} r_c$ corresponding to the distance $\sqrt{2} r_c$ from the cylindrical domain wall; $R_{cr}$ is the minimal radius of cylindrical domain originated from the **finite size effect**.

Electric potential reaches maximal value

$$\varphi_{max} \approx f_{12} \frac{(Q_{11} + Q_{12}) P_S^2}{(s_{11} + s_{12})} \left(1 - \frac{R_{cr}}{R}\right) \tag{15}$$

in the center of domain for small domains (e.g. nanodomains). Note, that the expressions for $\tilde{P}_\perp^{max}$, $\tilde{E}_\perp^{max}$ and $\varphi_{max}$ differ from the expressions listed in Ref.[4] by the factor $\left(1 - \frac{R_{cr}}{R}\right)$ originated from the finite size effect. As anticipated, $\tilde{P}_{\uparrow\uparrow}$ is maximal in the center of cylindrical domain and zero at its boundary $\rho = R$.

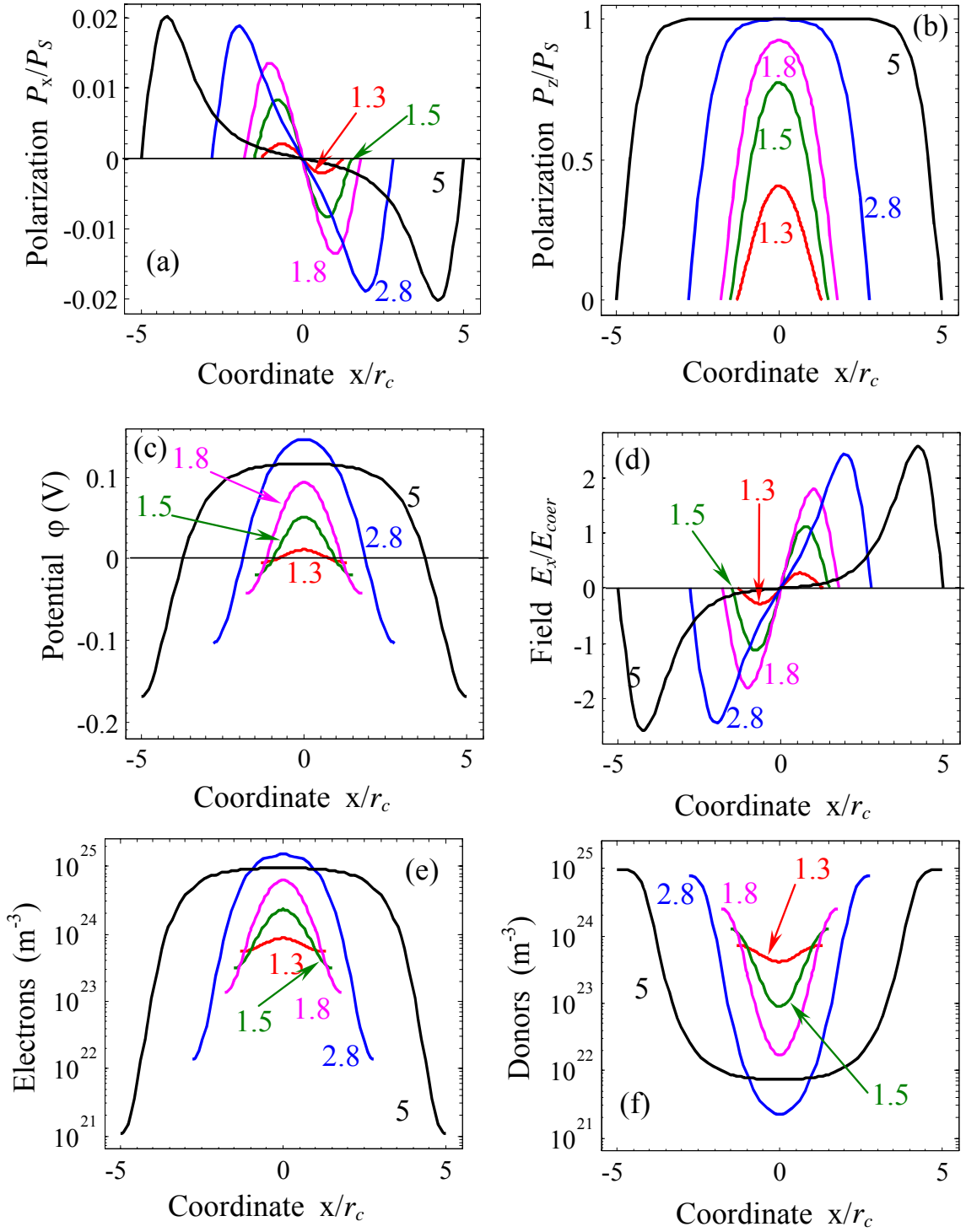

**Fig. 9.** Distributions of the polarization component $P_x(x)/P_S$ (perpendicular to the walls surface) (a) and (b) the component $P_z(x)/P_S$ parallel to the walls plane, potential $\varphi(x)$ (c), field $E_x(x)/E_{coer}$ (d), concentrations of electrons $n(x)$ (e) and ionized donors $N_d^+(x)$ (f) along the cross-section of cylindrical domain with different radius $R/r_c$=1.3, 1.5, 1.8, 2.8, 5 (shown near the curves) and **positive flexoelectric coupling coefficient** $f_{12} = +1 \times 10^{-10}$ m$^3$/C. Other parameters are same as for **Fig. 2**.

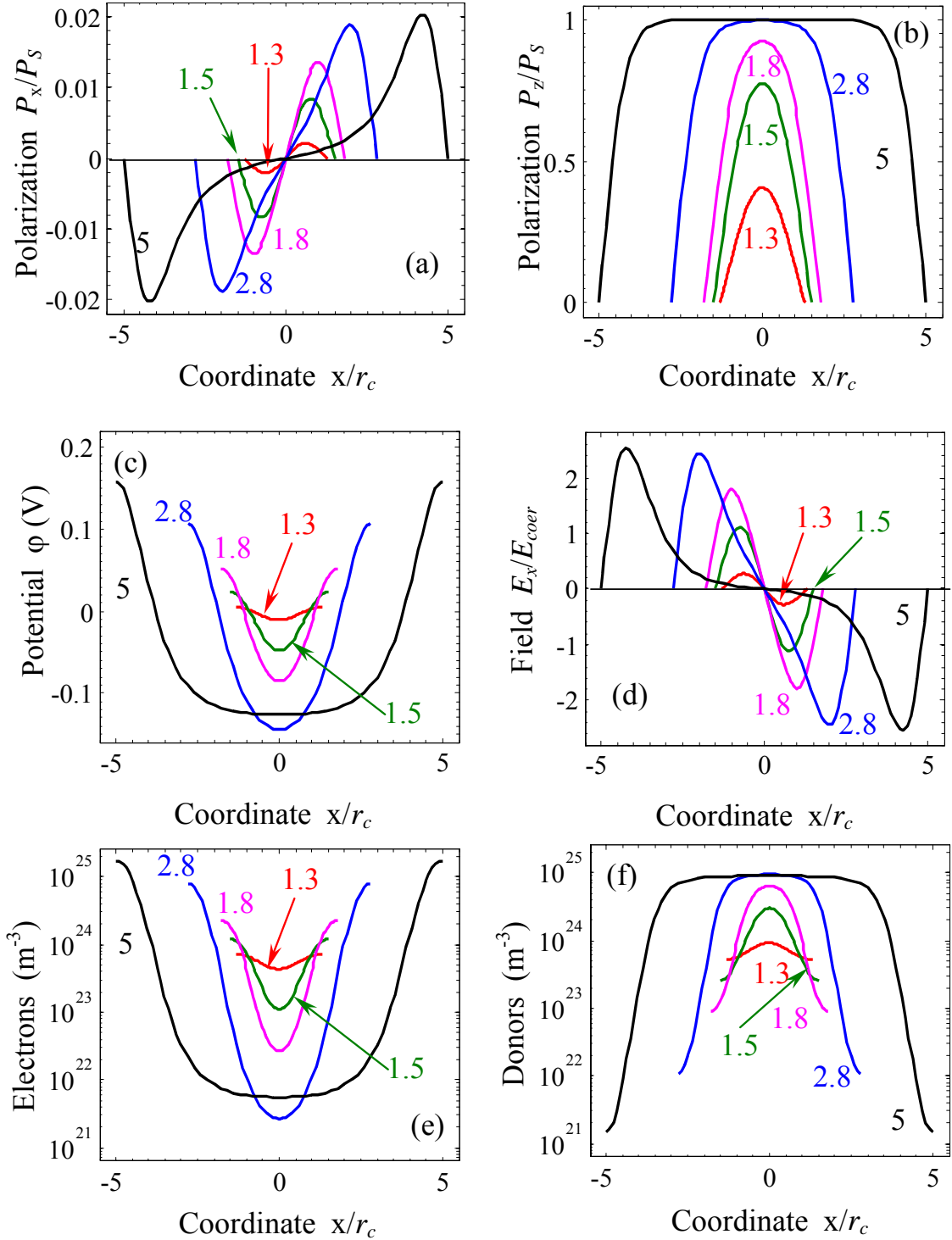

**Fig. 10.** Distributions of the polarization component $P_x(x)/P_S$ (perpendicular to the walls surface) (a) and (b) the component $P_z(x)/P_S$ parallel to the walls plane, potential $\varphi(x)$ (c), field $E_x(x)/E_{coer}$ (d), concentrations of electrons $n(x)$ (e) and ionized donors $N_d^+(x)$ (f) along the cross-section of cylindrical domain with different radius $R/r_c$=1.3, 1.5, 1.8, 2.8, 5 (shown near the curves) and **negative flexoelectric coupling coefficient** $f_{12} = -1 \times 10^{-10}$ m$^3$/C. Other parameters are same as for **Fig. 2**.

It is seen from the **Figs.9a-b** and **10**, that the decrease of the domain radius $R$ leads to the suppression of the polarization components maximum as well as to the decrease of the modulation depth of the potential and screening charges profiles along the domain cross-section. Polarization, potential, field and carries concentration profiles have sinusoidal shape for thin stripes. Deviation from the sinusoidal shape appears and strongly increases with $R$ increase.

It is seen from the **Figs.9e,f** and **10e,f**, that either electron or donor accumulation takes place in the nanodomain depending on the $f_{12}$ sign and spontaneous polarization direction. In contrast to thick domain stripes and thicker cylindrical domains, in which the carrier accumulation (and so the static conductivity) sharply increases at the domain walls only, thin nanodomains of radius $R \leq 5r_c$ can be conducting through entire their cross-section.

The size effects of the electron and donor accumulation/depletion by cylindrical domain walls and entire nanodomains are demonstrated in **Fig. 11a,b** for positive, zero and negative flexoelectric coupling coefficient $f_{12}$. To estimate the observable conductivity, concentrations of electrons $n(\rho)$ and donors $N_d^+(\rho)$ were averaged across the domain wall via distance $\rho \in \{R - r_c, R\}$ (solid curves) as well as entire the domain cross-section $\rho \in \{0, R\}$ (dashed curves).

Similarly to the case of domain stripes, the electronic conductivity of cylindrical domain wall monotonically increases and then saturates (up to 30 times in saturation in comparison with a bulk electronic conductivity) with the nanodomain radius increase for **negative** flexoelectric coupling (see solid curves in **Fig. 11a**). The ionic conductivity of cylindrical domain wall monotonically increases and then saturates (up to 20 times in saturation in comparison with a bulk ionic conductivity) with nanodomain radius increase for **positive** flexoelectric coupling (see solid curves in **Fig. 11b**).

Similarly to the case of domain stripe, the averaging entire the domain cross-section smears the impact of flexoelectric coupling sign, but dashed curves in **Figs.11a,b** firstly increase with radius $R$ increase for small domains radii $R_{cr} < R < 5r_c$, then reaches a maximum at $R \sim 5r_c$ and then decreases with further $R$ increase. The situation with electrons and ionized donors accumulation in the vicinity of cylindrical domain and in the central region of the domain is similar to the one discussed in the previous subsection for the case of domain stripes. Briefly free electrons accumulate in the vicinity of cylindrical domain walls for negative $f_{12}$, the central region of the domain is depleted with electrons (see **Fig.10e**). For positive $f_{12}$ the immediate vicinity of cylindrical domain walls are depleted with electrons, while the central region of the domain accumulates electrons (see **Fig.9e**). Ionized donors are accumulated in the immediate vicinity of cylindrical domain walls for positive $f_{12}$, while the central region of the domain is depleted with ionized donors (see **Fig.9f and 10f**). Due to the flexoelectric coupling even in average the cylindrical nanodomains of radius $R \sim 5r_c$ can be more conductive (up to 5 times) then the monodomain region with $\sigma_{e,i}(\infty)$.

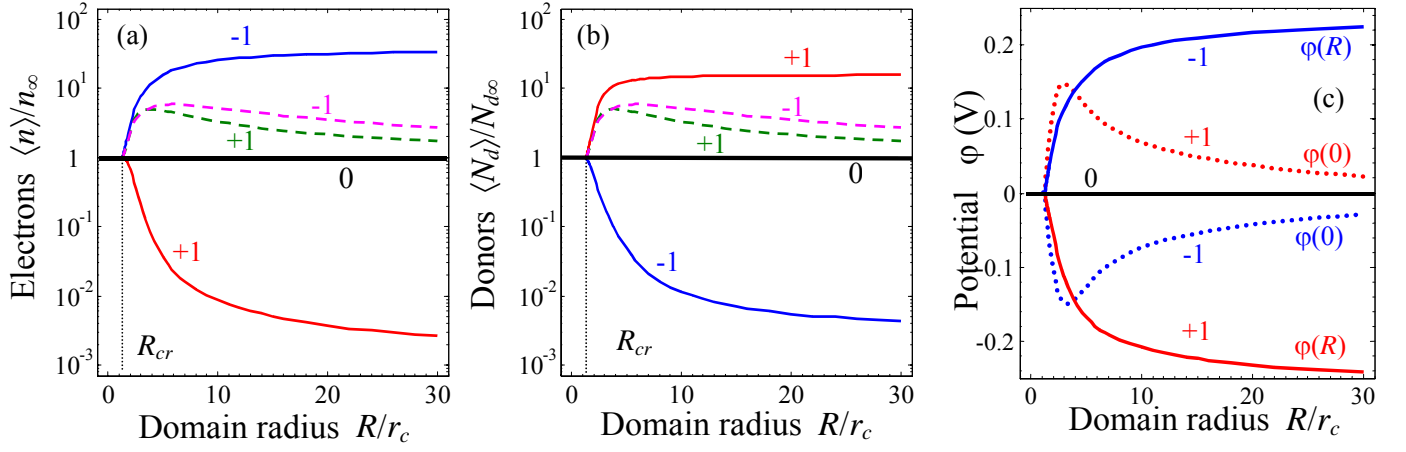

**Fig. 11.** Dependence of the relative electron $\langle n(\xi)\rangle/n(\infty)$ (a) and donor $\langle N_d^+(\xi)\rangle/N_d^+(\infty)$ (b) concentrations, potential $\varphi$ (c) on the radius $R$ of the cylindrical domain calculated for different **flexoelectric coupling coefficient** $f_{12}=-1\times10^{-10}$m$^3$/C, 0 and $f_{12}=-1\times10^{-10}$m$^3$/C (numbers near the curves). Concentrations of electrons $n(\rho)$ and donors $N_d^+(\rho)$ were averaged across the domain distance $\rho\in\{R-r_c, R\}$ (solid curves) as well as entire the domain cross-section $\rho\in\{0, R\}$ (dashed curves). Potential barrier (c) is plotted at the cylindrical domain wall ($\varphi(R)$, solid curves) and in the middle of the domain ($\varphi(0)$, dotted curves). Material parameters are the same as in **Figs.2.**

The carriers accumulation in the domain wall region is caused by the potential barrier $\varphi(\rho)$ in turn caused by the uncompensated bound charge $P_\perp(\rho)$ given by Eq.(13). Corresponding potential barrier is plotted in **Fig. 11c** at the curved domain wall ($\varphi(R)$, solid curves) and in the centre of the domain ($\varphi(0)$, dotted curves) for positive, zero and negative flexoelectric coupling coefficients. Potential barrier $\varphi(R)$ monotonically increases with the domain radius $R$ increase and then saturates. Potential $\varphi(0)$ in the centre of the domain firstly increases with $R$ increase, reaches maximum at $R\sim 3r_c$ and then decreases with further $R$ increase. For sub-micro and microdomains with radius $R\gg 100r_c$ the potential vanishes in the central region of the domain, i.e. $\varphi(\rho\ll R)\approx 0$, as anticipated.

## 4. Discussion and summary

Using LGD theory allowing for inhomogeneous elastic strains, flexoelectric coupling and electrostriction we performed analyses of the carriers accumulation across 180-degree domain wall in multi-axial *n*-type ferroelectric-semiconductors with donors and free electrons. Along with coupled LGD equations for the polarization components perpendicular and parallel to the wall plane, we solved the Poisson equation for electrostatic potential. Spatial distributions of the ionized shallow donors (e.g.

intrinsic oxygen vacancies), free electrons and holes were found self-consistently using the effective mass approximation for their energy density of states. Performed theoretical analyses shows that we meet with several scenarios of the 180-degree domain wall conduction in ferroelectric stripe and cylindrical domains, the choice between which depend on the wall geometry (tilt angle and domain shape and size), wall type (head-to-head or tail-to-tail) and the sign and value of the flexoelectric coupling coefficient. In particular

1) Similarly to the case of uniaxial *n*-type ferroelectric-semiconductors [15], the tilted wall is charged in the multi-axial ferroelectric-semiconductors and hence the electric field of the bound charge attracts free electrons and repulses ionized donors from the head-to-head wall region. The electron concentration is the highest when the wall plane is perpendicular to the spontaneous polarization direction at the wall (perpendicular domain wall); it decreases with the bound charge decrease and reaches minimum for the parallel domain wall. The carriers accumulation leads to the strong increase of the static conductivity across the charged domain walls in multi-axial ferroelectric-semiconductors up 3 orders of magnitude for the perpendicular domain walls in $Pb(Zr,Ti)O_3$. In contrast to uniaxial ferroelectrics, components perpendicular to the wall plane originate inside the wall region in the multi-axial ferroelectric-semiconductors.

2) The flexoelectric coupling, which is rather high for ferroelectric perovskites [41-44], leads to the appearance of polarization components perpendicular to the wall plane and its strong gradient across the wall even for nominally uncharged walls. Note, that the effect was visible in the first principle calculations (see Fig.12 in Ref.[52]). At the same time, the polarization component parallel to the wall plane is indifferent to the presence of the flexoelectric coupling and electrostriction coupling induces the narrowing of the domain wall**.** The polarization components perpendicular to the wall plane is directly related to the bound charge, in turn leading to the strong electric field wall and then to the free screening carriers accumulation across the wall. The carrier accumulation effect across the uncharged straight and cylindrical walls appeared to be significant, up to 10-30 times increase for domain stripes and cylindrical nanodomains in $Pb(Zr,Ti)O_3$ for the typical range of flexoelectric coefficients. The charge of accumulated carriers (electrons or donors for considered case) is determined by the sigh of the flexoelectric coefficient: negative coefficient leads to the accumulation of negative carriers (electrons or acceptors), positive coefficient leads to the accumulation of positive carriers (holes, donors or vacancies).

Note, that the static conductivity increase takes place in the p-type ferroelectric-semiconductors across the tail-to-tail walls (compare the **Tables 1a** and **1b**).

**Table 1a. Wall conductivity in the n-type ferroelectric-semiconductors**

| flexoelectric | Conductivity of | Conductivity of tilted head-to- | Conductivity of tilted |
|---|---|---|---|

| coupling coefficient $f_{12}$ | parallel 180-degree domain walls | head 180-degree walls | tail-to-tail 180-degree walls |
|---|---|---|---|
| positive | slightly higher than the bulk one (2-5 times) due to the donors accumulation | strongly increases with the tilt angle increase up to $10^3$ due to the electron accumulation by the bound charge | slightly higher than the bulk one (2-5 times) due to the donor and holes accumulation |
| zero | the same as in the bulk | strongly increases up to $10^3$ with the tilt angle increase due to the electron accumulation | slightly higher than the bulk one (2-5 times) due to the donor and holes accumulation |
| negative | increases up to 10-50 due to the electrons accumulation | strongly increases up to $10^3$ with the tilt angle increase due to the electron accumulation | slightly higher than the bulk one (2-5 times) due to the donor and holes accumulation |

**Table 1b. Wall conductivity in the p-type ferroelectric-semiconductors**

| flexoelectric coupling coefficient $f_{12}$ | Conductivity of parallel 180-degree domain walls | Conductivity of tilted head-to-head 180-degree walls | Conductivity of tilted tail-to-tail 180-degree walls |
|---|---|---|---|
| positive | increases up to 10-50 due to the holes accumulation | slightly higher than the bulk one (2-5 times) due to the acceptors and electrons accumulation | strongly increases with the bound charge increase up to $10^3$ due to the holes accumulation |
| zero | the same as in the bulk | slightly higher than the bulk one (2-5 times) due to the acceptors and electrons accumulation | strongly increases with the bound charge increase up to $10^3$ due to the holes accumulation |
| negative | slightly higher than the bulk one (2-5 times) due to the acceptors (or electrons) accumulation | slightly higher than the bulk one (2-5 times) due to the acceptors and electrons accumulation | strongly increases with the bound charge increase up to $10^3$ due to the holes accumulation |

3) The size effect of the electron and donor accumulation/depletion by thin stripe domains and cylindrical nanodomains is revealed. In contrast to thick domain stripes and thicker cylindrical domains, in which the carrier accumulation (and so the static conductivity) sharply increases at the domain walls only, thin nanodomains of radius less then 5-10 correlation length appeared conducting entire their cross-section. Such conductive nanosized channels may be promising for nanoelectronics development due to the possibility to control their spatial location by external stimulus (e.g. by nanomanipulation with the charged probe).

## Acknowledgements

SVK research was supported by the US Department of Energy Office of Basic Energy Sciences. EEA, ANM and GSS research was sponsored by Ukraine State Agency on Science, Innovation and Informatization (grants UU30/004 and GP/F32/099) as budget funds. ANM, EEA and GSS also acknowledge National Science Foundation (DMR-0908718) and user agreement with CNMS N UR-08-869.

# Supplementary materials

## Appendix A.

### A.1. Evident form of equations of state

The evident form of the flexoelectric coupling contribution (3c) is

$$\Delta G_{flexo} = -\frac{f_{11}}{2}\left(X_1\frac{\partial P_1}{\partial x_1} - P_1\frac{\partial X_1}{\partial x_1} + X_2\frac{\partial P_2}{\partial x_2} - P_2\frac{\partial X_2}{\partial x_2} + X_3\frac{\partial P_3}{\partial x_3} - P_3\frac{\partial X_3}{\partial x_3}\right) -$$

$$-\frac{f_{12}}{2}\begin{pmatrix} X_2\frac{\partial P_1}{\partial x_1} - P_1\frac{\partial X_2}{\partial x_1} + X_1\frac{\partial P_2}{\partial x_2} - P_2\frac{\partial X_1}{\partial x_2} \\ + X_1\frac{\partial P_3}{\partial x_3} - P_3\frac{\partial X_1}{\partial x_3} + X_3\frac{\partial P_1}{\partial x_1} - P_1\frac{\partial X_3}{\partial x_1} \\ + X_2\frac{\partial P_3}{\partial x_3} - P_3\frac{\partial X_2}{\partial x_3} + X_3\frac{\partial P_2}{\partial x_2} - P_2\frac{\partial X_3}{\partial x_2} \end{pmatrix} - \frac{f_{44}}{2}\begin{pmatrix} X_4\frac{\partial P_3}{\partial x_2} - P_3\frac{\partial X_4}{\partial x_2} + X_4\frac{\partial P_2}{\partial x_3} - P_2\frac{\partial X_4}{\partial x_3} \\ + X_5\frac{\partial P_1}{\partial x_3} - P_1\frac{\partial X_5}{\partial x_3} + X_5\frac{\partial P_3}{\partial x_1} - P_3\frac{\partial X_5}{\partial x_1} \\ + X_6\frac{\partial P_2}{\partial x_1} - P_2\frac{\partial X_6}{\partial x_1} + X_6\frac{\partial P_1}{\partial x_2} - P_1\frac{\partial X_6}{\partial x_2} \end{pmatrix} \quad (A.1)$$

Contribution of the inhomogeneous stresses to the free energy could be evaluated as :

$$\Delta F = \Delta G_{elast} + \Delta G_{strict} + \Delta G_{flexo} + \widetilde{X}_i\widetilde{u}_i =$$

$$= \frac{\left((Q_{11}+Q_{12})\left(\widetilde{P}_2^2 - (\widetilde{P}_2^{(S)})^2 + \widetilde{P}_3^2 - (\widetilde{P}_3^{(S)})^2\right) - (-2Q_{12})\left(\widetilde{P}_1^2 - (\widetilde{P}_1^{(S)})^2\right)\right)^2}{4(s_{11}+s_{12})} + \frac{f_{12}^2}{s_{11}+s_{12}}\left(\frac{\partial \widetilde{P}_1}{\partial \widetilde{x}_1}\right)^2 -$$

$$+ f_{12}\frac{\partial \widetilde{P}_1}{\partial \widetilde{x}_1}\frac{\left((Q_{11}+Q_{12})\left(\widetilde{P}_2^2 - (\widetilde{P}_2^{(S)})^2 + \widetilde{P}_3^2 - (\widetilde{P}_3^{(S)})^2\right) - (-2Q_{12})\left(\widetilde{P}_1^2 - (\widetilde{P}_1^{(S)})^2\right)\right)}{(s_{11}+s_{12})} + \quad (A.2)$$

$$+ \frac{(Q_{11}-Q_{12})^2}{4(s_{11}-s_{12})}\begin{pmatrix} \left((\widetilde{P}_2^{(S)})^2 - \widetilde{P}_2^2\right)^2 + \left((\widetilde{P}_3^{(S)})^2 - \widetilde{P}_3^2\right)^2 + \\ + 2\left((\widetilde{P}_2^{(S)})^2 + \widetilde{P}_2^2\right)\left((\widetilde{P}_3^{(S)})^2 + \widetilde{P}_3^2\right) - 8\widetilde{P}_2^{(S)}\widetilde{P}_3^{(S)}\widetilde{P}_2\widetilde{P}_3 \end{pmatrix}$$

Here we employ isotropy approximation for elastic compliances, electrostriction coefficients:

$$s_{44} = 2(s_{11} - s_{12}), \quad Q_{44} = 2(Q_{11} - Q_{12}), \quad f_{44} = f_{11} - f_{12} \quad (A.3)$$

For derivation of Eq.(A.2) see **Appendix C**.

Evident form of other contributions to the free energy (1) could be obtained by either coordinate system transformation of by substituting $\Delta G_b[P_i] \equiv \Delta G_b[(A^T)_{ij}\widetilde{P}_j]$. Thus, equations determining distributions of $\widetilde{P}_j$ could be written as

$$\left.\frac{\partial \Delta G_b[\mathbf{P}]}{\partial P_i}\right|_{\mathbf{P}\to\hat{A}^T\widetilde{\mathbf{P}}}(A^T)_{i1} - \frac{\left((Q_{11}+Q_{12})\left(\widetilde{P}_2^2 - (\widetilde{P}_2^{(S)})^2 + \widetilde{P}_3^2 - (\widetilde{P}_3^{(S)})^2\right) - (-2Q_{12})\left(\widetilde{P}_1^2 - (\widetilde{P}_1^{(S)})^2\right)\right)}{(s_{11}+s_{12})}(-2Q_{12})\widetilde{P}_1 +$$

$$- f_{12}\frac{\partial}{\partial \widetilde{x}_1}\left(\frac{\left((Q_{11}+Q_{12})\left(\widetilde{P}_2^2 - (\widetilde{P}_2^{(S)})^2 + \widetilde{P}_3^2 - (\widetilde{P}_3^{(S)})^2\right) - (-2Q_{12})\left(\widetilde{P}_1^2 - (\widetilde{P}_1^{(S)})^2\right)\right)}{(s_{11}+s_{12})}\right) - \left(g_{11} + \frac{2f_{12}^2}{s_{11}+s_{12}}\right)\frac{\partial^2 \widetilde{P}_1}{\partial \widetilde{x}_1^2} = \widetilde{E}_1$$

(A.4a)

$$\left. \frac{\partial \Delta G_b[\mathbf{P}]}{\partial P_i} \right|_{\mathbf{P} \to \hat{A}^T \tilde{\mathbf{P}}} (A^T)_{i2} + \frac{\left((Q_{11}+Q_{12})\left(\tilde{P}_2^2 - (\tilde{P}_2^{(S)})^2 + \tilde{P}_3^2 - (\tilde{P}_3^{(S)})^2\right) - (-2Q_{12})\left(\tilde{P}_1^2 - (\tilde{P}_1^{(S)})^2\right)\right)}{(s_{11}+s_{12})}(Q_{11}+Q_{12})\tilde{P}_2 +$$

$$+ \frac{(Q_{11}-Q_{12})^2}{(s_{11}-s_{12})}\left(\tilde{P}_2\left((\tilde{P}_3^{(S)})^2 - (\tilde{P}_2^{(S)})^2 + \tilde{P}_3^2 + \tilde{P}_2^2\right) - 2\tilde{P}_2^{(S)}\tilde{P}_3^{(S)}\tilde{P}_3\right) - g_{44}\frac{\partial^2 \tilde{P}_2}{\partial \tilde{x}_1^2} = \tilde{E}_2$$

(A.4b)

$$\left. \frac{\partial \Delta G_b[\mathbf{P}]}{\partial P_i} \right|_{\mathbf{P} \to \hat{A}^T \tilde{\mathbf{P}}} (A^T)_{i3} + \frac{\left((Q_{11}+Q_{12})\left(\tilde{P}_2^2 - (\tilde{P}_2^{(S)})^2 + \tilde{P}_3^2 - (\tilde{P}_3^{(S)})^2\right) - (-2Q_{12})\left(\tilde{P}_1^2 - (\tilde{P}_1^{(S)})^2\right)\right)}{(s_{11}+s_{12})}(Q_{11}+Q_{12})\tilde{P}_3 +$$

$$+ \frac{(Q_{11}-Q_{12})^2}{(s_{11}-s_{12})}\left(\tilde{P}_3\left(-(\tilde{P}_3^{(S)})^2 + (\tilde{P}_2^{(S)})^2 + \tilde{P}_3^2 + \tilde{P}_2^2\right) - 2\tilde{P}_2^{(S)}\tilde{P}_3^{(S)}\tilde{P}_2\right) - g_{44}\frac{\partial^2 \tilde{P}_3}{\partial \tilde{x}_1^2} = \tilde{E}_3$$

(A.4c)

Here $P_S^2 \equiv (\tilde{P}_1^{(S)})^2 + (\tilde{P}_2^{(S)})^2 + (\tilde{P}_3^{(S)})^2 \equiv (P_1^{(S)})^2 + (P_2^{(S)})^2 + (P_3^{(S)})^2$.

*A.2. System of equations in dimensionless variables*

Let us introduce the dimensionless variables:

$$p_i = \tilde{P}_j / P_S, \quad p_j^{(S)} = \tilde{P}_j^{(S)} / P_S, \quad \Phi = q\varphi / k_B T, \quad x = \frac{\tilde{x}_1}{r_c}, \quad \tilde{E}_1 = -\frac{\partial \varphi}{\partial \tilde{x}_1} = -\frac{\partial \Phi}{\partial x}\frac{k_B T}{r_c q} \quad (A.5a)$$

where dimensionless electric field is $-\frac{\partial \Phi}{\partial x}$ and the correlation length

$$r_c = \sqrt{g_{44}/(-2a_1)} \quad (A.5b)$$

Dimensionless parameters, characterizing electrostriction contribution

$$q_{11} = \frac{(Q_{11}+Q_{12})P_S^2}{\sqrt{-2a_1 P_S^2 (s_{11}+s_{12})}}, \quad q_{12} = \frac{(-2Q_{12})P_S^2}{\sqrt{-2a_1 P_S^2 (s_{11}+s_{12})}}, \quad q_{44} = \frac{(Q_{11}-Q_{12})P_S^2}{\sqrt{-2a_1 P_S^2 (s_{11}-s_{12})}} \quad (A.5c)$$

Using Eqs.(A.5), Eqs.(A.4) becomes

$$\frac{(A^T)_{i1} G_{b,i}}{-2a_1 P_S} - \left(q_{11}\left(p_2^2 - (p_2^{(S)})^2 + p_3^2 - (p_3^{(S)})^2\right) - q_{12}\left(p_1^2 - (p_1^{(S)})^2\right)\right)q_{12} p_1 +$$

$$- \frac{2f_{12}}{\sqrt{g_{44}(s_{11}+s_{12})}}\left(q_{11}\left(p_2\frac{\partial p_2}{\partial x} + p_3\frac{\partial p_3}{\partial x}\right) - q_{12} p_1\frac{\partial p_1}{\partial x}\right) - $$

$$-\left(\frac{g_{11}}{g_{44}} + \frac{2f_{12}^2}{(s_{11}+s_{12})g_{44}}\right)\frac{\partial^2 p_1}{\partial x^2} + \frac{k_B T}{(-2a_1 P_S)r_c q}\frac{\partial \Phi}{\partial x} = 0$$

(A.6a)

$$\frac{(A^T)_{i2} G_{b,i}}{-2a_1 P_S} + \left(q_{11}\left(p_2^2 - (p_2^{(S)})^2 + p_3^2 - (p_3^{(S)})^2\right) - q_{12}\left(p_1^2 - (p_1^{(S)})^2\right)\right)q_{11} p_2 +$$

$$+ q_{44}^2\left(p_2\left((p_3^{(S)})^2 - (p_2^{(S)})^2 + p_3^2 + p_2^2\right) - 2p_2^{(S)}p_3^{(S)}p_3\right) - \frac{\partial^2 p_2}{\partial x^2} = 0$$

(A.6b)

$$\frac{(A^T)_{i3}G_{b,i}}{-2a_1P_S} + \left(q_{11}\left(p_2^2 - (p_2^{(S)})^2 + p_3^2 - (p_3^{(S)})^2\right) - q_{12}\left(p_1^2 - (p_1^{(S)})^2\right)\right)q_{11}p_3 +$$
$$+ q_{44}^2\left(p_3\left(-(p_3^{(S)})^2 + (p_2^{(S)})^2 + p_3^2 + p_2^2\right) - 2p_2^{(S)}p_3^{(S)}p_2\right) - \frac{\partial^2 p_3}{\partial x^2} = 0 \quad (A.6c)$$

Here we took into account that in the coordinate system electric field has only one component, $\tilde{E}_1$. Equation for electrostatic potential:

$$\frac{\partial^2 \Phi}{\partial x^2} = \frac{(-2a_1P_S)r_c q}{(-2a_1\varepsilon_0\varepsilon_b)k_B T}\frac{\partial p_1}{\partial x} - \left(\frac{N_d^+ + p - n - N_a^-}{N_{d0}}\right)\frac{r_c^2 N_{d0} q^2}{\varepsilon_0\varepsilon_b k_B T} \quad (A.7)$$

## Appendix B. Material parameters

**Table B.1.** Free energy parameters for bulk ferroelectric PbZr$_{1-x}$Ti$_x$O$_3$ (from Refs.[53, 54])

| x | 0.4 | 0.6 | 0.8 | 1 |
|---|---|---|---|---|
| $a_1$ (10$^7$C$^{-2}$·m$^2$N) at 25°C | -7.904 | -8.34 | -14.84 | -17.08 |
| $a_{11}$(10$^8$C$^{-4}$·m$^6$N) | 1.362 | 0.3614 | -0.305 | -0.73 |
| $a_{12}$(10$^8$C$^{-4}$·m$^6$N) | 2.391 | 3.233 | 6.32 | 7.5 |
| $a_{111}$(10$^8$C$^{-6}$·m$^{10}$N) | 2.713 | 1.859 | 2.475 | 2.61 |
| $a_{112}$(10$^8$C$^{-6}$·m$^{10}$N) | 12.13 | 8.503 | 9.684 | 6.1 |
| $a_{123}$(10$^8$C$^{-6}$·m$^{10}$N) | -56.9 | -40.63 | -49.01 | -36.6 |
| $Q_{11}$(C$^{-2}$·m$^4$) | 0.0726 | 0.0812 | 0.0814 | 0.089 |
| $Q_{12}$(C$^{-2}$·m$^4$) | -0.0271 | -0.0295 | -0.0245 | -0.026 |
| $Q_{44}$(C$^{-2}$·m$^4$) | 0.0629 | 0.0671 | 0.0642 | 0.0675 |
| $s_{11}$ (in units of 10$^{-12}$ Pa$^{-1}$) | 8.8 | 8.6 | 8.2 | 8.0 |
| $s_{12}$ (in units of 10$^{-12}$ Pa$^{-1}$) | -2.9 | -2.8 | -2.6 | -2.5 |
| $s_{44}$ (in units of 10$^{-12}$ Pa$^{-1}$) | 24.6 | 21.2 | 14.4 | 9.0 |

in particular the coefficient $f_{12} \approx -(0.5-1)\cdot 10^{-10}$m$^3$/C is involved in our calculations, gradient coefficients $g_{11}=2.0\cdot 10^{-10}$C$^{-2}$m$^4$N, $g_{44}=1.0\cdot 10^{-10}$C$^{-2}$m$^4$N, correlation length $r_c = \sqrt{-g_{44}/2a_1} \approx 0.5$ nm, $m_n = 0.05m_e$, $m_p = 5m_e$, where $m_e$ is the mass of the free electron, $\varepsilon_{33}^b = 5$, band gap $E_g = 3$ eV, $N_{d0} = 10^{25}$ m$^{-3}$, $E_d = 0.1$eV.

## Appendix C. Mechanical equilibrium at 180°-domain wall

Elastic equation of state could be obtained by the minimization $\delta G/\delta X_{jk} = -u_{jk}$. Subscripts 1, 2 and 3 denote Cartesian coordinates *x, y, z* and Voigt's (matrix) notations are used. We assume that bulk material paraelectric phase has cubic symmetry (e.g. $Q_{11} = Q_{22} = Q_{33}$ etc). Below we employ isotropy approximation for elastic compliances, electrostriction coefficients, i.e.

$$s_{44} = 2(s_{11} - s_{12}), \quad Q_{44} = 2(Q_{11} - Q_{12}), \quad f_{44} = f_{11} - f_{12} \tag{C.1}$$

Equations of state, including the flexoelectric contributions, have the form:

$$\tilde{u}_1 - u_1^{(F)} = s_{11}\tilde{X}_1 + s_{12}(\tilde{X}_2 + \tilde{X}_3), \tag{C.2a}$$

$$\tilde{u}_2 - u_2^{(F)} = s_{11}\tilde{X}_2 + s_{12}(\tilde{X}_1 + \tilde{X}_3), \tag{C.2b}$$

$$\tilde{u}_3 - u_3^{(F)} = s_{11}\tilde{X}_3 + s_{12}(\tilde{X}_1 + \tilde{X}_2), \tag{C.2c}$$

$$\tilde{u}_4 - u_4^{(F)} = s_{44}\tilde{X}_4, \tag{C.2d}$$

$$\tilde{u}_5 - u_5^{(F)} = s_{44}\tilde{X}_5, \tag{C.2d}$$

$$\tilde{u}_6 - u_6^{(F)} = s_{44}\tilde{X}_6. \tag{C.2e}$$

Here we introduced designations for the strains in the absence of any elastic stresses:

$$u_1^{(F)} = f_{11}\frac{\partial \tilde{P}_1}{\partial x_1} + Q_{11}\tilde{P}_1^2 + Q_{12}(\tilde{P}_2^2 + \tilde{P}_3^2), \tag{C.3a}$$

$$u_2^{(F)} = f_{12}\frac{\partial \tilde{P}_1}{\partial \tilde{x}_1} + Q_{11}\tilde{P}_2^2 + Q_{12}(\tilde{P}_1^2 + \tilde{P}_3^2), \tag{C.3b}$$

$$u_3^{(F)} = f_{12}\frac{\partial \tilde{P}_1}{\partial \tilde{x}_1} + Q_{11}\tilde{P}_3^2 + Q_{12}(\tilde{P}_2^2 + \tilde{P}_1^2), \tag{C.3c}$$

$$u_4^{(F)} = Q_{44}\tilde{P}_2\tilde{P}_3, \quad u_5^{(F)} = f_{44}\frac{\partial \tilde{P}_3}{\partial \tilde{x}_1} + Q_{44}\tilde{P}_1\tilde{P}_3, \quad u_6^{(F)} = f_{44}\frac{\partial \tilde{P}_2}{\partial \tilde{x}_1} + Q_{44}\tilde{P}_1\tilde{P}_2. \tag{C.3d}$$

Note, that elastic contribution to the free energy could be written as

$$\Delta G_{elast} + \Delta G_{strict} + \Delta G_{flexo} = \begin{pmatrix} -\frac{1}{2}s_{11}(\tilde{X}_1^2 + \tilde{X}_2^2 + \tilde{X}_3^2) - s_{12}(\tilde{X}_1\tilde{X}_2 + \tilde{X}_2\tilde{X}_3 + \tilde{X}_3\tilde{X}_1) \\ -\frac{1}{2}s_{44}(\tilde{X}_4^2 + \tilde{X}_5^2 + \tilde{X}_6^2) - \\ -\tilde{X}_1 u_1^{(F)} - \tilde{X}_2 u_2^{(F)} - \tilde{X}_3 u_3^{(F)} - \tilde{X}_4 u_4^{(F)} - \tilde{X}_5 u_5^{(F)} - \tilde{X}_6 u_6^{(F)} \end{pmatrix} \tag{C.4}$$

Here we used integration in parts for flexoelectric contribution (A.1) to be rewritten as:

$$\Delta G_{flexo} = -f_{11}\left(X_1\frac{\partial P_1}{\partial x_1} + X_2\frac{\partial P_2}{\partial x_2} + X_3\frac{\partial P_3}{\partial x_3}\right) -$$

$$- f_{12}\left(X_2\frac{\partial P_1}{\partial x_1} + X_1\frac{\partial P_2}{\partial x_2} + X_1\frac{\partial P_3}{\partial x_3} + X_3\frac{\partial P_1}{\partial x_1} + X_2\frac{\partial P_3}{\partial x_3} + X_3\frac{\partial P_2}{\partial x_2}\right) - \tag{C.5}$$

$$- f_{44}\left(X_4\frac{\partial P_3}{\partial x_2} + X_4\frac{\partial P_2}{\partial x_3} + X_5\frac{\partial P_1}{\partial x_3} + X_5\frac{\partial P_3}{\partial x_1} + X_6\frac{\partial P_2}{\partial x_1} + X_6\frac{\partial P_1}{\partial x_2}\right)$$

Compatibility relation $e_{ikl}e_{jmn}(\partial^2 \tilde{u}_{ln}/\partial \tilde{x}_k \partial \tilde{x}_m) = 0$ leads to the conditions of constant strains $\tilde{u}_2 = const$, $\tilde{u}_3 = const$, $\tilde{u}_4 = const$, while general form dependences like $\tilde{u}_1 = \tilde{u}_1(\tilde{x}_1)$, $u_5 = u_6(x_1)$ and $\tilde{u}_6 = \tilde{u}_6(\tilde{x}_1)$ do not contradict to these relations. Mechanical equilibrium conditions $\partial \tilde{X}_{ij}/\partial \tilde{x}_i = 0$ lead to $\partial \tilde{X}_1/\partial \tilde{x}_1 = 0$, $\partial \tilde{X}_5/\partial \tilde{x}_1 = 0$, $\partial \tilde{X}_6/\partial \tilde{x}_1 = 0$. Since $\tilde{X}_{ij}(\tilde{x}_1 \to \pm\infty) = 0$, one obtains $\tilde{X}_1 = \tilde{X}_5 = \tilde{X}_6 = 0$. In this case some of the shear strains are trivial, $\tilde{u}_5 = u_5^{(F)}$, $\tilde{u}_6 = u_6^{(F)}$.

Setting constant strains $u_2$, $u_3$ and $u_4$ to values of spontaneous strains in stress free, homogeneous system,

$$\tilde{u}_2 = u_2^{(S)} \equiv Q_{11}\left(\tilde{P}_2^{(S)}\right)^2 + Q_{12}\left(\left(\tilde{P}_3^{(S)}\right)^2 + \left(\tilde{P}_1^{(S)}\right)^2\right), \tag{C.6a}$$

$$\tilde{u}_3 = u_3^{(S)} \equiv Q_{11}\left(\tilde{P}_3^{(S)}\right)^2 + Q_{12}\left(\left(\tilde{P}_2^{(S)}\right)^2 + \left(\tilde{P}_1^{(S)}\right)^2\right), \tag{C.6b}$$

$$\tilde{u}_4 = u_4^{(S)} \equiv Q_{44}\tilde{P}_2^{(S)}\tilde{P}_3^{(S)}. \tag{C.6c}$$

one can rewrite the system of equations (C.2) with respect to unknown stresses ($\tilde{X}_2, \tilde{X}_3$).

$$s_{11}\tilde{X}_2 + s_{12}\tilde{X}_3 = u_2^{(S)} - u_2^{(F)} \tag{C.7a}$$

$$s_{12}\tilde{X}_2 + s_{11}\tilde{X}_3 = u_3^{(S)} - u_3^{(F)} \tag{C.7b}$$

while strain $u_1$ could be expressed via $X_2$ and $X_3$. Finally, inhomogeneous stress and strain components can be written in the form:

$$\tilde{X}_2(\tilde{x}_1) = \frac{s_{11}\left(u_2^{(S)} - u_2^{(F)}\right) - s_{12}\left(u_3^{(S)} - u_3^{(F)}\right)}{s_{11}^2 - s_{12}^2}, \tag{C.8a}$$

$$\tilde{X}_3(\tilde{x}_1) = \frac{s_{11}\left(u_3^{(S)} - u_3^{(F)}\right) - s_{12}\left(u_2^{(S)} - u_2^{(F)}\right)}{s_{11}^2 - s_{12}^2}, \tag{C.8b}$$

$$\tilde{X}_4(\tilde{x}_1) = \frac{u_4^{(S)} - u_4^{(F)}}{s_{44}}, \tag{C.8c}$$

$$\tilde{u}_1 = u_1^{(F)} + \frac{s_{12}}{s_{11}+s_{12}}\left(u_2^{(S)} - u_2^{(F)} + u_3^{(S)} - u_3^{(F)}\right). \tag{C.8d}$$

For the cases of the partially clamped system (such as domain walls) one should find the equilibrium state as the minimum of the Helmholtz free energy $F = G + \int_V d^3r \cdot u_{jk} X_{jk}$ originating from Legendre transformation of $G$. Thus, contribution of the inhomogeneous stresses to the free energy could be evaluated as

$$\Delta F = \Delta G + \tilde{X}_i \tilde{u}_i = \Delta G_{elast} + \Delta G_{strict} + \Delta G_{flexo} + \tilde{X}_i \tilde{u}_i =$$
$$= -\frac{1}{2}s_{11}\left(\tilde{X}_2^2 + \tilde{X}_3^2\right) - s_{12}\tilde{X}_2\tilde{X}_3 - \frac{1}{2}s_{44}\tilde{X}_4^2 + \tilde{X}_2\left(\tilde{u}_2 - u_2^{(F)}\right) + \tilde{X}_3\left(\tilde{u}_3 - u_3^{(F)}\right) + \tilde{X}_4\left(\tilde{u}_4 - u_4^{(F)}\right) = \tag{C.9}$$
$$= \frac{\left(\left(u_2^{(S)} - u_2^{(F)}\right) + \left(u_3^{(S)} - u_3^{(F)}\right)\right)^2}{4(s_{11}+s_{12})} + \frac{\left(\left(u_2^{(S)} - u_2^{(F)}\right) - \left(u_3^{(S)} - u_3^{(F)}\right)\right)^2}{4(s_{11}-s_{12})} + \frac{\left(u_4^{(S)} - u_4^{(F)}\right)^2}{2s_{44}}$$

Using the evident form of strains we derived that:

$$\Delta F = \frac{\left((Q_{11}+Q_{12})\left(\left(\widetilde{P}_2^{(S)}\right)^2 - \widetilde{P}_2^2 + \left(\widetilde{P}_3^{(S)}\right)^2 - \widetilde{P}_3^2\right) + 2Q_{12}\left(\left(\widetilde{P}_1^{(S)}\right)^2 - \widetilde{P}_1^2\right)\right)}{4(s_{11}+s_{12})} + \frac{f_{12}^2}{s_{11}+s_{12}}\left(\frac{\partial \widetilde{P}_1}{\partial \widetilde{x}_1}\right)^2 -$$

$$- f_{12}\frac{\partial \widetilde{P}_1}{\partial \widetilde{x}_1}\frac{\left((Q_{11}+Q_{12})\left(\left(\widetilde{P}_2^{(S)}\right)^2 - \widetilde{P}_2^2 + \left(\widetilde{P}_3^{(S)}\right)^2 - \widetilde{P}_3^2\right) + 2Q_{12}\left(\left(\widetilde{P}_1^{(S)}\right)^2 - \widetilde{P}_1^2\right)\right)}{(s_{11}+s_{12})} + \quad (C.10)$$

$$+ \frac{(Q_{11}-Q_{12})^2}{4(s_{11}-s_{12})}\left(\begin{array}{l}\left(\left(\widetilde{P}_2^{(S)}\right)^2 - \widetilde{P}_2^2\right)^2 + \left(\left(\widetilde{P}_3^{(S)}\right)^2 - \widetilde{P}_3^2\right)^2 + \\ + 2\left(\left(\widetilde{P}_2^{(S)}\right)^2 + \widetilde{P}_2^2\right)\left(\left(\widetilde{P}_3^{(S)}\right)^2 + \widetilde{P}_3^2\right) - 8\widetilde{P}_2^{(S)}\widetilde{P}_3^{(S)}\widetilde{P}_2\widetilde{P}_3\end{array}\right)$$

**Appendix D. Nominally neutral 180-degree domain wall.**

Considering nominally neutral 180-degree wall with wall normal along $\widetilde{x}_1 \equiv x_1$ axis we could use pseudocubic crystallographic frame, and rewrite equations determining distributions of polarization components as

$$2P_1(a_1 + a_{12}P_3^2 + a_{112}P_3^4) + 4(a_{11} + a_{112}P_3^2)P_1^3 + 6a_{111}P_1^5$$
$$+ \frac{((Q_{11}+Q_{12})(P_3^2 - (P_S)^2) + 2Q_{12}P_1^2)}{(s_{11}+s_{12})}(2Q_{12})P_1 + \quad (D.1a)$$
$$- f_{12}\frac{\partial}{\partial x_1}\left(\frac{((Q_{11}+Q_{12})(P_3^2 - (P_S)^2) + 2Q_{12}P_1^2)}{(s_{11}+s_{12})}\right) - \left(g_{11} + \frac{2f_{12}^2}{s_{11}+s_{12}}\right)\frac{\partial^2 P_1}{\partial x_1^2} = E_1$$

$$2P_3(a_1 + a_{12}P_1^2 + a_{112}P_1^4) + 4(a_{11} + a_{112}P_1^2)P_3^3 + 6a_{111}P_3^5 +$$
$$+ \frac{2Q_{12}(Q_{11}+Q_{12})P_3P_1^2}{(s_{11}+s_{12})} + \left(\frac{(Q_{11}-Q_{12})^2}{(s_{11}-s_{12})} + \frac{(Q_{11}+Q_{12})^2}{(s_{11}+s_{12})}\right)P_3\left(P_3^2 - \left(P_3^{(S)}\right)^2\right) - g_{44}\frac{\partial^2 P_3}{\partial x_1^2} = 0 \quad (D.1b)$$

Here we consider only two components of polarization, since far from the wall we have only the one component $P_3 = \pm P_S$) and normal to the wall component, $P_1$, could be induced by flexoelectric coupling.

Supposing that $P_1 \ll P_3$, we could liberalize (D.1) with respect to $P_1$ and get the following equations

$$2P_1\left(a_1 + a_{12}P_3^2 + a_{112}P_3^4 + \frac{Q_{12}(Q_{11}+Q_{12})}{(s_{11}+s_{12})}\left(P_3^2 - (P_S)^2\right)\right) -$$
$$- f_{12}\frac{\partial}{\partial x_1}\left(\frac{(Q_{11}+Q_{12})}{(s_{11}+s_{12})}P_3^2\right) - \left(g_{11} + \frac{2f_{12}^2}{s_{11}+s_{12}}\right)\frac{\partial^2 P_1}{\partial x^2} \approx E_1 \quad (D.2a)$$

$$2a_1P_3 + 4a_{11}P_3^3 + 6a_{111}P_3^5 + \left(\frac{(Q_{11}-Q_{12})^2}{(s_{11}-s_{12})} + \frac{(Q_{11}+Q_{12})^2}{(s_{11}+s_{12})}\right)P_3\left(P_3^2 - (P_S)^2\right) - g_{44}\frac{\partial^2 P_3}{\partial x_1^2} = 0 \quad (D.2b)$$

Below we introduce electrostatic potential, $\mathbf{E}^d = -\nabla\varphi$, which should satisfy the Poisson equation

$$\frac{\partial^2 \varphi}{\partial x_1^2} \approx \frac{1}{\varepsilon_0\varepsilon_b}\frac{\partial P_1}{\partial x_1} + \frac{\varphi}{R_d^2} \quad (D.3)$$

Here we suppose that changes of potential are small, so we could use Debye approximation and introduce screening radius $R_d$.

Note, that for the numerical solution of the equations (D.1)-(D.3) we used the following boundary conditions at the center of domain $E_1 = 0$, $P_1 = 0$, $\frac{\partial P_3}{\partial x_1} = 0$ and $E_1 = 0$, $P_1 = 0$, $P_3 = 0$ at the domain wall.

One could see that now Eq. (D.2a) could be solved independently (this is corroborated by the results of numerical calculations) and then the solution for $P_3(x_1)$ could be inserted into Eq.(D.2b) as inhomogeneity:

$$\alpha P_1 - g \frac{\partial^2 P_1}{\partial x_1^2} + \frac{\partial \varphi}{\partial x_1} = f_{12} \frac{(Q_{11} + Q_{12})}{(s_{11} + s_{12})} \frac{\partial P_3^2}{\partial x_1} \tag{D.4a}$$

Here we introduced designations

$$\alpha \equiv 2\left(a_1 + a_{12}P_3^2 + a_{112}P_3^4 + \frac{Q_{12}(Q_{11} + Q_{12})}{(s_{11} + s_{12})}\left(P_3^2 - (P_3^{(S)})^2\right)\right), \tag{D.4b}$$

$$g \equiv g_{11} + \frac{2f_{12}^2}{s_{11} + s_{12}}. \tag{D.4c}$$

Below we suppose that $\alpha \approx const$. In order to get closed form solution we neglect Debye screening and using boundary conditions for potential $\varphi(r \to \infty) = 0$, $\varphi(r \to -\infty) = 0$. This allows us to neglect the term $\frac{\varphi}{R_d^2}$ in Eq.(D.3) get the simplified version for (D.3)

$$\frac{\partial \varphi}{\partial x_1} \approx \frac{P_1}{\varepsilon_0 \varepsilon_b} \tag{D.5a}$$

and Eq.(D.4a)

$$\left(\alpha + \frac{1}{\varepsilon_0 \varepsilon_b}\right) P_1 - g \frac{\partial^2 P_1}{\partial x_1^2} = f_{12} \frac{(Q_{11} + Q_{12})}{(s_{11} + s_{12})} \frac{\partial P_3^2}{\partial x_1} \tag{D.5b}$$

Since we consider isolated domain wall we could use the boundary conditions like $P_1(x_1 \to \infty) = 0$, $P_1(x_1 \to -\infty) = 0$, we finally get the solution of (D.5b) (see e.g. [55])

$$P_1(x_1) = \frac{f_{12}\varepsilon_0\varepsilon_b}{\alpha\varepsilon_0\varepsilon_b + 1} \frac{(Q_{11} + Q_{12})}{(s_{11} + s_{12})} \int_{-\infty}^{\infty} P_3(x)\frac{\partial P_3(x)}{\partial x} l_d \exp\left(-\frac{|x - x_1|}{l_d}\right) dx \tag{D.6}$$

Here we introduced the length scale $l_d = \sqrt{\frac{g\varepsilon_0\varepsilon_b}{\alpha\varepsilon_0\varepsilon_b + 1}}$. Note, that even in $P^4$-approximation (when $P_3(x) = P_S \tanh(x/w)$) the integral (D.6) could be evaluated only numerically.

Below we employ the approximation of "splay step" like profile

$$P_3 \approx P_S \begin{cases} -1, & x < -w; \\ x/w, & -w \leq x \leq w; \\ 1, & x > w. \end{cases} \quad (D.7)$$

Here w is an effective half width of the wall (or polarization gradient at the wall center is $P_S/w$).

Then using Eq.(D.7) and integration over parts in Eq.(D.6), we derived an analytic expression:

$$P_1(x_1) = \frac{f_{12}\varepsilon_0\varepsilon_b}{\alpha\varepsilon_0\varepsilon_b + 1} \frac{(Q_{11}+Q_{12})P_S^2}{(s_{11}+s_{12})w} \begin{cases} 2\exp\left(\frac{x_1}{l_d}\right)\left(-\frac{l_d}{w}\sinh\left(\frac{w}{l_d}\right)+\cosh\left(\frac{w}{l_d}\right)\right), & x < -w; \\ 2\left(\frac{x_1}{w} - \exp\left(-\frac{w}{l_d}\right)\left(\frac{l_d}{w}+1\right)\sinh\left(\frac{x_1}{l_d}\right)\right), & -w \leq x \leq w; \\ 2\exp\left(-\frac{x_1}{l_d}\right)\left(-\frac{l_d}{w}\sinh\left(\frac{w}{l_d}\right)+\cosh\left(\frac{w}{l_d}\right)\right), & x > w. \end{cases} \quad (D.8a)$$

The extreme values of (D.8a) are reached near the points $x \approx \pm w$:

$$P_1(\pm w) = \pm \frac{f_{12}\varepsilon_0\varepsilon_b}{\alpha\varepsilon_0\varepsilon_b + 1} \frac{(Q_{11}+Q_{12})P_S^2}{(s_{11}+s_{12})w}\left(1-\frac{l_d}{w}+\exp\left(-2\frac{w}{l_d}\right)\left(1+\frac{l_d}{w}\right)\right) \approx$$
$$\approx \pm f_{12}\varepsilon_0\varepsilon_b \frac{(Q_{11}+Q_{12})P_S^2}{(s_{11}+s_{12})w}\left(1-\frac{l_d}{w}\right) \quad (D.8b)$$

Here we also employ that in the most cases $\alpha\varepsilon_0\varepsilon_b \ll 1$ and $w \gg l_d$.

Using well-known parameters for PbTiO$_3$ and $f_{12} = 10^{-10}$, $\varepsilon_b = 5$, w=1 nm, $l_d \sim 0.1$ nm one could get maximal value of $P_1$ about 0.02 C/m$^2$ (2 μC/cm$^2$), which is in excellent agreement with *ab initio* calculation of Behera et al [52].

Our calculations have shown that numerical solution of system of Eqs.(D.2)-(D.3) showed is almost independent on Debye screening radius for $R_d >$1 nm and agrees well with estimation (D.8b).

The depth of the potential well/height of the barrier on the wall due to $P_1(x_1)$-effect could be estimated from (D.5a) and (D.8a) as $\delta\varphi \approx f_{12}\frac{(Q_{11}+Q_{12})P_S^2}{(s_{11}+s_{12})}$. It is about 0.3 eV for PbTiO$_3$ with $f_{12} = 10^{-10}$ m$^3$/C.